\documentclass[journal=jacsat,manuscript=article]{achemso}
\usepackage[version=3]{mhchem} % Formula subscripts using \ce{}
\usepackage[T1]{fontenc}       % Use modern font encodings
\usepackage{lmodern}
\usepackage{relsize,exscale}
\usepackage{color}
\usepackage[dvinames]{xcolor}
\usepackage{pifont,color}
\usepackage{tikz}

\SectionNumbersOn
\setkeys{acs}{keywords = true}
\setcitestyle{super}
\setkeys{acs}{articletitle = true}

\newcommand*{\citen}[1]{%
  \begingroup
    \romannumeral-`\x % remove space at the beginning of \setcitestyle
    \setcitestyle{numbers}%
    \cite{#1}%
  \endgroup   
}

\author{Mohammadhasan Dinpajooh}
\affiliation{Department of Chemistry and Biochemistry, and Institute of Theoretical Science, University of Oregon, Eugene, Oregon 97403}
\author{Marina G. Guenza}
\affiliation{Department of Chemistry and Biochemistry, and Institute of Theoretical Science, University of Oregon, Eugene, Oregon 97403}
\email{mguenza@uoregon.edu}

\title{Coarse-Graining Simulation Approaches for Polymer Melts: Range of Potential and Computational Efficiency}

\begin{document}

\begin{abstract}
The integral equation coarse-graining (IECG) approach is a promising high-level coarse-graining (CG) method
for polymer melts, with variable resolution from soft spheres to multi CG sites, 
which preserves the structural and thermodynamical consistencies with the related atomistic simulations. 
When compared to the atomistic description, the procedure of coarse-graining results in smoother free energy surfaces, 
longer-ranged potentials, a decrease in the number of interaction sites for a given polymer, and more.
Because these changes have competing effects on the computational efficiency of the CG model,
care needs to be taken when studying the effect of coarse-graining on the computational speed-up
in CG molecular dynamics simulations. For instance, treatment of long-range CG interactions requires the selection of cutoff distances that include the attractive part of the effective CG potential and force. 
In particular, we show how the complex nature of the range and curvature of the effective CG potential,
the selection of a suitable CG timestep, the choice of the cutoff distance, the molecular dynamics algorithms, and
the smoothness of the CG free energy surface affect the efficiency of IECG simulations. 
By direct comparison with the atomistic simulations of relatively short chain polymer melts,
we find that the overall computational efficiency is highest for the highest level of CG (soft spheres),
with an overall improvement of the computational efficiency  being about
$10^6-10^8$ for various CG levels/resolutions.
Therefore, the IECG method can have important 
applications in molecular dynamics simulations of polymeric systems.  
Finally, making use of the standard spatial decomposition algorithm, the parallel scalability 
of the IECG simulations for various levels of CG is presented.
Optimal parallel scaling is observed for a reasonably large number of processors.
\end{abstract}

\section{Introduction}
\label{introsec}
The properties of polymeric liquids develop on such an extended range of time- and lengthscales that 
they cannot be directly explored by standard atomistic Molecular Dynamics (MD) simulations.
While MD simulations are useful in providing a microscopic picture of the mechanisms that determine 
the macroscopic properties measured experimentally,\cite{Rahman1971,Allen1987,Frenkel2002} 
the computational efficiency of atomistic simulations for polymeric liquids still doesn't allow the 
direct comparison with experimental data.\cite{Salerno2016}
It has been shown that by simplifying the molecular description, thus reducing the degrees of freedom, MD simulations are able to afford a considerable computational speed-up, 
opening the way to extended simulation studies in time- and lengthscales never covered before.\cite{Reith2003,Harmandaris2006,Salerno2016,Guenza2018} This computational speed-up is a consequence of the many effects of this coarse-graining (CG) procedure on the potential, which are complex and sometimes competing.
To the best of our knowledge, a comprehensive and detailed analysis of the effects of coarse-graining on the computational efficiency of CG-MD has not yet been performed. We aim at addressing this point through a careful study of the effects of coarse-graining on the computational efficiency of CG-MD simulations, starting from the Integral Equation Coarse-Graining (IECG) method.\cite{Guenza2018,Clark2013,Dinpajooh2017}  
The IECG-MD simulations have the advantage of predicting, among other properties, quantitatively consistent radial distribution function and pressure with the related atomistic simulations.\cite{Clark2012,McCarty2014,Dinpajooh2017a,Dinpajooh2017}  Furthermore, the IECG potential has been analytically solved for liquids of long polymer chains, and low granularity coarse-grained description.\cite{Yatsenko2004,Clark2010,Clark2012,Clark2013,McCarty2012} Density-dependent studies of the IECG method have proved transferability of the potential through mapping on the Carnahan-Starling equation of state.\cite{McCarty2014,Dinpajooh2017,Carnahan1970} 

Extensive CG methods have been developed in recent years.\cite{Carbone2008,Harmandaris2003,Das2012,Ramos2015,Cao2015}  
In general, a CG procedure involves averaging a number of local details into an effective CG site through a mapping scheme.
This gives an effective CG potential that is a free energy of the system and as such, is state dependent.
Even at the thermodynamic sate of calibration, most CG models fail to simultaneously 
reproduce all properties accurately, which is known as the representability problem.
In addition, the CG parameters optimized for a set of
thermodynamic states do not usually apply to other thermodynamic conditions,
which is known as the transferability problem.
It is often the case that a trial function is used in the CG simulation and 
the parameters in this function are optimized numerically to reproduce a target atomistic property. 
The target property can be defined in atomistic simulations that need to be performed at the start, 
in the so-called bottom-up approaches,
\cite{Harmandaris2006,Izvekov2005,Shell2008,Brini2013,Dunn2015,Rudzinski2015,Salerno2016} or 
from experimental data in top-down CG models.\cite{Avendano2011,Gil-Villegas1997}
Hybrid approaches, which combine both bottom-up and top-down strategies have also been developed.\cite{Hsu2015}  

The IECG method approaches coarse-graining from a first-principles perspective, 
as this formalism is rooted on liquid-state theory and on the solution of the
Ornstein-Zernike equation.\cite{Yatsenko2004,Clark2012,Clark2013,McCarty2014,Dinpajooh2017,Guenza2018}
This model has the advantage of depending only on one non-trivial parameter,
the direct correlation function at zero wavevector ($c_0$),
which can be determined either from a bottom-up or a top-down procedure.
This direct correlation function follows the proper equation of state making the
IECG potential transferable.
In this work, we focus on the bottom-up IECG approach to assess and optimize 
its computational efficiency; therefore, the IECG simulations are compared with the
underlying atomistic simulations in the same thermodynamic conditions.  
While analytical solution of structural and thermodynamical properties, as well as IECG MD simulations have shown quantitative agreement, within numerical error for IECG-MD, with the atomistic MD simulations for radial distribution functions, equation of state, and free energy,\cite{McCarty2014,Dinpajooh2017,Dinpajooh2017a}
in this study, we specifically focus on the pair correlation functions and on the pressure,
for which direct comparison of atomistic and IECG simulations from a bottom-up approach 
perspective is a reasonable test. 

As a consequence of coarse-graining, a number of monomeric sites 
becomes represented by a single CG interaction site. 
Choosing the granularity of this description depends on the amount of 
molecular details the CG model should retain to 
investigate the physical problem of interest.\cite{Kremer1990,Harmandaris2006,McCarty2014,Dinpajooh2017} 
The range of the effective CG potential depends on the extent of CG. 
The analytical solution of the IECG potential for polymer melts 
shows that the range of the effective CG potential, expressed in units of the CG size, scales with the number of 
monomers inside the CG unit (level of CG) as a power of $1/4$.\cite{Clark2012}
One should expect that fine-graining models are relatively more long-ranged than
the atomistic ones\cite{Izvekov2005,Salerno2016} but less long-ranged than the effective CG potential in 
the IECG method, which involves coarse-graining to a higher degree.\cite{Clark2012}
However, care must be taken in constructing short-ranged CG potentials for highly coarse-grained systems, 
because the range and curvature of the effective CG potential can have complex effects on the 
efficiency of various levels of CG and the accuracy of the physical properties measured in the CG-MD simulation. 

In addition, the fastest motions of the CG system are substantially slower in their characteristic correlation times 
than the atomistic bond fluctuations, thus making possible the use of much larger timesteps in the CG-MD simulations 
than in the atomistic simulations. 
More importantly, the CG potential becomes softer and longer-ranged as the extent of CG becomes larger.
The range of the effective CG potential being significantly larger than the atomistic one
leads to important implications in the computational efficiency,
as we discuss in detail in this paper. 
In general, averaging the related degrees of freedom in the CG process
results in a relatively smoother CG free energy surface, with
the dynamics of the CG system significantly enhanced.\cite{Fritz2009,Lyubimov2010,Lyubimov2011,Fritz2011,Lyubimov2013,Davtyan2016}
Therefore, the computational efficiency for a given CG approach may be determined 
by considering the relative number of pairwise interactions and timestep
for atomistic and CG simulations, as well as the dynamical enhancement in CG simulations, 
as quantified by eq \ref {Eqeff} and discussed in Section \ref{results}.

One final aspect that we will address in this paper is
the relevance of the MD algorithms on the efficiency of 
the IECG method.\cite{Allen1987,Frenkel2002,Plimpton1995,Bussi2007,Halverson2013} 
Since the seminal paper by Verlet,\cite{Verlet1967} the Verlet neighbor list has been used
extensively to reduce the computational time in MD simulations. 
Chialvo and Debenedetti\cite{Chialvo1990} performed a thorough study of the optimum neighbor list
radius and update frequency for atomic Lennard-Jones systems and linear rigid triatomics at various 
timesteps, system sizes, temperatures, and densities. Their simulation results showed that 
the optimum neighbor list radius increases with temperature and MD timestep and decreases with density.
They reported optimum neighbor list radii ranging from $0.1$ to $0.5$ $\sigma$ at various state points, 
where $\sigma$ is the intermolecular separation for which the potential energy is zero. 
The more complex nature of the CG interactions than the ones for simple atomistic systems indicates the need to investigate the optimum values of the neighborhood list radii, which is required in performing IECG-MD simulations.

Although the comparison with atomistic simulations is reasonable when studying 
many structural and thermodynamical properties via molecular simulations,
in the case of phase transitions\cite{Schweizer1997,Stanley1987}, where
finite size effects\cite{Stukan2002} are
extremely important, atomistic simulations become problematic.
While the IECG method can explore more extended regions of
time and space,\cite{McCarty2010} in those conditions
a direct test of CG methods against atomistic simulations would not be 
possible and an evaluation of the computational speed-up is not feasible.

Finally, while we notice that the IECG is a theory specific for CG of polymeric liquids,
the effects of coarse-graining on the efficiency of CG-MD simulations presented 
here are general and should hold qualitatively, albeit not quantitatively,
for any CG model of molecular liquids.

The paper is organized as follows: 
In Section \ref{theoryp} the IECG theory is briefly discussed to define the properties relevant to this study, 
while Section \ref{simtot} illustrates the procedure used in performing atomistic and IECG simulations. 
The calculation of the efficiency and the factors that determine its value are discussed in Section \ref{results}. 
This section consists of a number of subsections that contain: a brief introduction to the equation defining the
computational efficiency (Section \ref{definition}); 
the evaluation of the number of pairwise interactions and 
the role of the range of the effective CG potential in determining this number 
(Section \ref{Nprsec}); 
the determination of the proper CG timestep as a function of the level of CG (Section \ref{dtsec}); 
and the effects of the enhancement of the dynamics, due to coarse-graining, on computational efficiency
(Section \ref{alphasec}).
Section \ref{effsec} presents the overall computational efficiency of the IECG simulations
for various levels of CG, from the soft spheres to multi CG site models,
as directly compared with the related atomistic simulations. 
Finally, the parallel scalability of the IECG simulations are presented in the last section.
A brief discussion concludes the paper.

%%%%%%%%%%%%%%%%%%%%%%%%%%%%%%%%%%%%%%%%%%%%%%%%%%%%%

\section{Theoretical Background}
\label{theoryp}

The IECG method is a theoretical approach to coarse-grain polymer liquids, based on the solution of the Ornstein-Zernike integral equation theory.\cite{Hansen2003} IECG is an implementation in the coarse-graining framework of the PRISM approach\cite{Schweizer1987,Schweizer1990} by Schweizer and Curro, which is an extension to polymeric liquids of the RISM approach by Chandler and Andersen.\cite{Chandler1972} 
While atomistic MD simulations provide useful information in the temporal trajectory of the space and momentum coordinates,
from which all the structural, thermodynamics, and dynamic quantities can be calculated,
they are limited to the range of length- and timescales that they can cover when simulating polymeric liquids, 
and rarely reach the conditions typically sampled experimentally. 
On the other hand, the PRISM approach has no significant limitations in the range of lengthscales that it can cover,
and has been validated for a wide number of polymeric systems,\cite{Guenza1997,Yatsenko2004,Yatsenko2005,Sankar2015} 
but it provides information only at the level of the pair distribution functions, and their Fourier transforms, 
analogous to a scattering experiment.
Thus, the IECG method addresses a much needed niche in the study of complex polymeric liquids,
because it connects the flexibility and accuracy of the PRISM theory to the
convenient representation of the temporal evolution of the system through space coordinate 
and velocity trajectories.
Furthermore, the IECG-MD simulations have been shown to produce pair distribution functions, 
pressure, equation of state, and 
excess free energies in quantitative agreement with the
atomistic simulations,\cite{Clark2013,McCarty2014,Dinpajooh2017,Dinpajooh2017a}
while further extending the time- and lengthscales that can be simulated.
Thus, the IECG method is an ideal complementary technique to atomistic simulations of polymeric systems. 

\begin{figure}[htb]
\includegraphics[width=8cm]{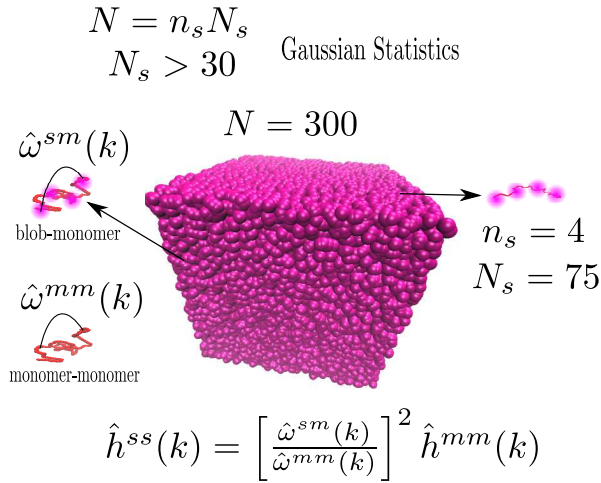}
\caption{ 
A snapshot of IECG simulations consisting of polymer chains represented as four CG sites ($n_s=4$),
where each polymer chain consists of $300$ monomers ($N=300$).
Note that the CG correlation functions can be obtained in terms of monomer correlation
functions (see Eq. \ref{hss}), and the number of monomers in each CG site is large enough 
to allow one to use the Gaussian statistics to obtain the intramolecular correlations.}
\label{iecgfig}
\end{figure}

In the IECG model, each polymer is represented by one or more CG sites, 
(see Figure \ref{iecgfig}), where each site consists of a large enough number of monomers
that Gaussian statistics applies for the intramolecular correlations (the lengthscale of a site has to be larger than the polymer persistence length).
Given a liquid of $n$ polymers in a volume $V$, the chain density is $\rho_{ch}=n/V$, while the monomer density is $\rho_m=\rho_{ch}/N$, with $N$ the number of monomers in a chain. When a chain is partitioned in $n_s$ sites, such that the number of monomers in a site is $N_s=N/n_s$, the site density is  $\rho_s=n n_s/V$. For $n_s=1$ each polymer is represented as a soft sphere. 
Considering the CG units as auxiliary sites, the total correlation function between the CG units for 
a system consisting of the monomeric and CG sites in reciprocal space, $\hat{h}^{ss}(k)$, is obtained as a combination of intramolecular, $\hat{\omega}(k)$, and monomeric total intermolecular, $\hat{h}^{mm}(k)$
correlation functions as
\begin{equation}
\hat{h}^{ss}(k) = \left[ \hat{\omega}^{sm}(k)/\hat{\omega}^{mm}(k) \right]^2 \hat{h}^{mm}(k),
\label{hss}
\end{equation}
where $s$ is the index of the CG unit and $m$ is the index of the monomeric (atomistic) unit.
Making use of the Ornstein-Zernike (OZ) equation and the PRISM approach,
the monomer total intermolecular correlation function, $\hat{h}^{mm}(k)$, is given in Fourier space by
\begin{equation}
\hat{h}^{mm} (k) = \hat{\omega}^{mm}(k) \hat{c}^{mm}(k) \left[  \hat{\omega}^{mm}(k) + \rho_m \hat{h}^{mm} (k)  \right] = \frac{ \hat{\omega}^{mm}(k)  \hat{c}^{mm}(k) \hat{\omega}^{mm}(k) }{ 1-  \rho_m \hat{c}^{mm}(k) \hat{\omega}^{mm}(k) },
\label{OZhmm}
\end{equation}
where  $\hat{c}^{mm}(k)$ is the monomer-monomer direct correlation function, which can be approximated by its behavior at large values of distance or small values of $k$ such as the limit of zero wavevector $k$, $c_0=\hat{c}^{mm}(k \rightarrow 0)$. The value of $c_0$ may be determined from the pressure values obtained from 
the atomistic simulations, $P_{\rm{AT}}$,  via 
\begin{equation}
c_0 = \frac{2}{N \rho_m} - \frac{2 P_{\rm{AT}}}{\rho_m^2 k_{\rm{B}} T },
\label{c0}
\end{equation}
where $k_{\rm{B}}$ is the Boltzmann constant, $T$ is the temperature, and the aforementioned equation of state 
is derived in previous works.\cite{Clark2012,Dinpajooh2017a,Dinpajooh2017} Therefore, making use of eq \ref{hss} one can get the 
total correlation function between CG units. The direct correlation functions between CG sites, $\hat{c}^{ss}(k)$,
are then determined by solving the OZ equation for a system consisting of the CG sites only, and
the effective CG potential is then generated using the appropriate closure, such as the HyperNetted Chain (HNC) closure:\cite{Louis2000} 
$U^{ss}(r)/(k_{\rm{B}}T)=- \ln{[h^{ss}(r)+1]}+h^{ss}(r)-c^{ss}(r)$ (also see our previous work for more details). 
\cite{Clark2010,Clark2012,Clark2013,McCarty2014,Dinpajooh2017}  
In addition, the IECG Python-based programs and usage instructions as well as related knowledgebase materials are presented on a dedicated documentation Web site \cite{IECGweb}. 

\begin{figure}[htb]
\includegraphics[width=12cm]{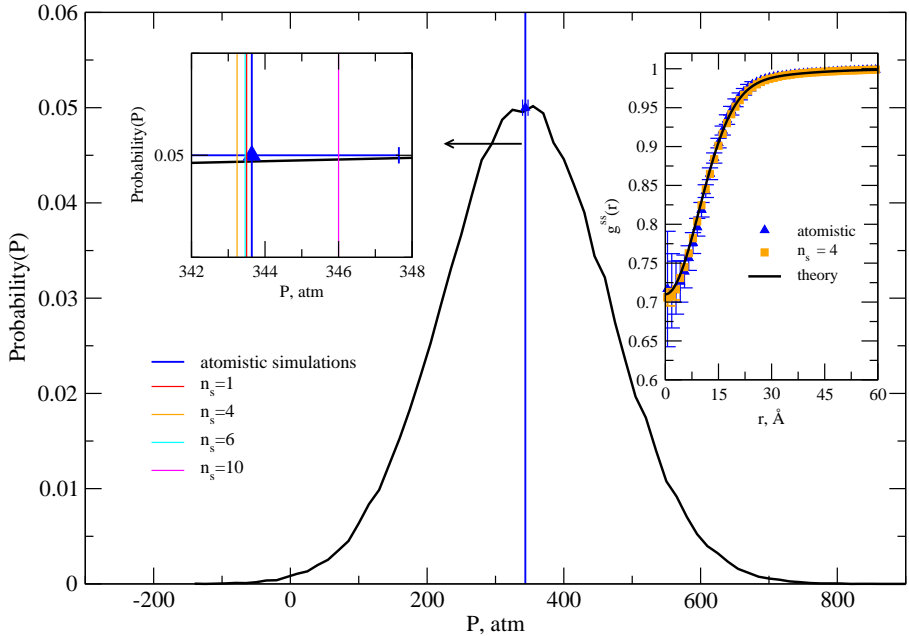}
\caption{
Structural and thermodynamical consistencies illustrated for a polymer melt with $N=300$, at
$503$ K and a monomer density of $0.03296$$ $\AA$^{-3}$.
Pressure distribution for the atomistic simulation (black curve).
The blue line shows the average pressure for atomistic simulations with the simulation error bars
obtained from block averages (see left panel).
The red, orange, cyan, and magenta lines show the average pressure for coarse-grained (CG) simulations
with $1$, $4$, $6$, and $10$ sites per chain ($n_s$), respectively.
Theoretical and simulated radial distribution functions (right panel) when the polymer is represented by four coarse-grained sites.
Atomistic simulations (blue triangle) are compared with coarse-grained simulations (red square) and
with the theoretical prediction (black line).
The theoretical prediction and the data from the coarse-grained simulation are both within the error of the atomistic simulation.
}
\label{RDFTcons}
\end{figure}

The IECG-MD simulations can be used as an efficient way to investigate large-scale molecular mechanisms of interest in polymeric systems.
The results obtained from the IECG simulations have been shown to be structurally and thermodynamically
consistent with the atomistic ones.\cite{Clark2013,McCarty2014,Dinpajooh2017,Dinpajooh2017a}
Figure \ref{RDFTcons} illustrates the structural and thermodynamical consistencies for
a polymer melt with a degree of polymerization of $300$ at a given state point.
Note that an incorrect construction of the potential by using incompatible intramolecular distributions and intermolecular closures can lead to the breaking of the correlation between the pair distribution function and the thermodynamic properties, with the consequent loss of consistency with atomistic properties.\cite{Dinpajooh2017a}

It follows, that once combined with the atomistic simulations in a multiscale modeling procedure,
the IECG approach is able to cover a wide range of lengthscales of interest.\cite{McCarty2009} 
Furthermore, by making use of appropriate equation of states the IECG method can be applied to other 
state points, which makes the method transferable.
\cite{Clark2010,Clark2012,Clark2013,McCarty2014,Dinpajooh2017} 

In the limit of a liquid of long-polymer chains, for which the distribution of CG sites in the polymer chain
follows the Markovian statistics, the IECG potential has been solved analytically,
providing a clear connection between the range of the potential and the physical properties of the system coarse-grained.
The solution shows that the leading term in the range of the potential normalized by $n_s^{1/2}$ 
increases with the number of monomers in the CG site as $N_s^{1/4}$, indicating that the range of the potential increases 
with increasing level of CG. 
This effect can have complex and competing consequences on the computational efficiency of the IECG simulations, 
as discussed in the upcoming sections.\cite{Clark2012}

\section{Simulation Details}
\label{simtot}

\subsection{Atomistic Simulation Details}
\label{sim}

The molecular dynamics (MD) software program LAMMPS\cite{Plimpton1995} was used for all simulations.
All simulations were performed in canonical NVT ensemble
with 3-dimensional boundary conditions, Nos$\rm{\acute{e}}$-Hoover thermostat,
and standard velocity-Verlet integrator.
The TraPPE united atom force field\cite{Martin1998},
which used a harmonic potential for adjacent intramolecular sites,
was used for the MD simulations.
See Table \ref{param} for more details.
A cutoff distance of $14$ \AA~ were used and both potential and force were
required to go smoothly to zero at the cutoff distance
by multiplying the potential by the Mei-Davenport-Fernando (MDF) taper function\cite{Mei1991}

\begin{equation}
t(x) =
\begin{cases}  
  1 & \mathrm{if} \; r \leq r_m \\
  (1-x)^3(1+3x+6x^2) & \mathrm{if} \;  r_m < r \leq r_c \quad, \\
  0 & \mathrm{if} \; r > r_c 
\end{cases}  
\label{taper}
\end{equation}
where $r_c$ is the cutoff distance and we used an $r_m$ value of $12$ \AA\ and $x$ is given by

\begin{equation}
x = \frac{r-r_m}{r_c - r_m}
\label{xmdf}
\end{equation}

\begin{table}[tbh]
\centering
\caption{TraPPE united atom force field used in this work.}
\begin{tabular}{ccc}
     \hline
     \multicolumn{2}{c}{Bond potential: $U_{\rm{bond}} =  k_b (l-l_0)^2$ }  \\
     \hline
     CH$_2$--CH$_2$  & $k_b$, kcal mol$^{-1}$ \AA$^{-2}$ & $l_0$, \AA \\
       & $450$ &  $1.54$ \\
     \hline
     \multicolumn{2}{c}{Angle potential: $U_{\rm{angle}} =  k_\theta (\theta-\theta_0)^2$ }  \\
     CH$_2$--CH$_2$--CH$_2$  & $k_\theta$, kcal mol$^{-1}$ rad$^{-2}$ & $\theta_0$, deg \\
       & $62.1$ & $114$ \\
     \hline
     \multicolumn{2}{c}{Dihedral potential: $U_{\rm{dih}} =  \Sigma_i^3 \frac{C_i}{2} \left( 1+ e_i {\rm{cos}} ( i \phi) \right)$ }  \\
     CH$_2$--CH$_2$--CH$_2$--CH$_2$  & $C_i$, kcal mol$^{-1}$ & $e_i$ \\
     $i=1$  & $1.4110$  & $+$ \\
     $i=2$  & $-0.2708$ & $-$ \\
     $i=3$  & $3.1430$   & $+$ \\
     \hline
     \multicolumn{2}{c}{Non-bonded potential: $U_{\rm{LJ}} =  4\epsilon \left[ (\sigma/r)^{12} - (\sigma/r)^6 \right] $ }  \\
     CH$_2$--CH$_2$--CH$_2$--CH$_2$  & $\epsilon$, kcal mol$^{-1}$ & $\sigma$, \AA \\
       & $0.0912$ & $3.95$ \\
     \hline
\end{tabular}
\label{param}
\end{table}

The atomistic simulations were performed for the polymers with
degrees of polymerizations, $N$, of $44$, $192$, and $300$
at a monomer density of $0.03296$ sites
\AA$^{-3}$ at $503.17$ K, where the system sizes consisting
of $350$, $350$, and $300$ polymer chains were used.
To investigate the effect of density
on the relevant aspects of the computational efficiency,
atomistic simulations were also performed at $503$ K at monomer
densities of $0.03201$, $0.0334$, and $0.03439$ \AA$^{-3}$.
Similarly, the atomistic simulations were also performed at a monomer
density of $0.03296$ \AA$^{-3}$ at $473$, $543$, and $563$ K to
investigate the temperature dependence.
For all atomistic MD simulations, polymer chains were randomly generated,
and overlapping chains in the
initial configuration were slowly pushed apart by a soft repulsive potential.
Next, the full non-bonded potential was switched
on with a small timestep, and the system was run for an additional
$1$ ns while ramping up the timestep to $1.25$ fs. Subsequently,
chains were allowed to equilibrate before
final production runs were used with a timestep of $2$ fs and
a neighbor skin distance of $2$ \AA.
The production periods consisted of $80$, $700$, and $215$ ns for the polymer
melts with $N$ of $44$, $192$, and $300$, respectively.
The square root of average square end-to-end distance, $\langle R^2 \rangle^{1/2}$,
from atomistic simulations for $N$ of $44$, $192$, and $300$
were obtained as $26.8$, $60.7$, and $76.4$ \AA, respectively.

\subsection{IECG Simulation Details}
\label{simsec}
Unless mentioned all IECG-MD simulations were performed with the following simulation protocol:
the canonical ensemble was used with the Nos$\rm{\acute{e}}$-Hoover thermostat 
and standard velocity-Verlet integrator.
Periodic boundary conditions
were applied in all three dimensions. 
Intra and inter molecular effective IECG potentials, as described 
in ref \citen{Dinpajooh2017} were adopted, noting that in the multi-site CG models
the nonbonded intrachain effective CG potential 
involves intramolecular CG sites that are separated more than two apart.  
Details about using various timesteps, cutoff distances ($r_{\rm{cut}}$), 
and neighbor skin distances ($r_{\rm{skin}}$) for different approaches 
are presented in the following as needed.
The combined Verlet neighbor list and the link-cell binning algorithm 
were used to build the Verlet neighbor list, which was updated every 
10 steps.\cite{Allen1987} 
This algorithm has been used to obtain practical CG timesteps for 
various levels of CG considering 
the curvature and range of the effective CG potential (see Section \ref{dtsec}).  
The spatial decomposition of simulation domain was used for studying 
the parallel scalability, where the Verlet neighbor list was updated every 
step and the parallel IECG simulations were performed, 
making use of the standard frequency of writing the output files 
and trajectory files, which include the positions of particles.
Note that in the parallel atomistic MD simulations the same standard frequency 
of the IECG-MD simulations was used for writing the output files and trajectory files.

%%%%%%%%%%%%%%%%%%%%%%%%%%%%%%%%%%%%%%%%%%%%%%%%%%%%%%%%%%%%
\section{Calculation of the IECG Simulation Efficiency}
\label{results}
\subsection{Definition of Efficiency}
\label{definition}

Given the discussions presented in the previous sections, 
the computational efficiency of a CG approach at a given density and temperature can be quantified 
by the following equation:   
\begin{equation}
\varepsilon(N, n_s) = N_{pr} \: \Delta t_r \: \alpha, 
\label{Eqeff}
\end{equation}
where $N_{pr}$ is the ratio of the total number of pairwise interactions in the atomistic 
simulations to the ones in the CG simulations,
$\Delta t_r$ is the ratio of the timesteps used in the CG simulations to 
the ones used in the atomistic simulations, 
and $\alpha$ is the dynamical scaling factor, which corresponds to the speed-up of the dynamics due to coarse-graining. 
Note that the CG particles can undergo large displacements due to the increase in the diffusion constants,
which has implications in the selection of the appropriate neighbor skin distance, $r_{\rm{skin}}$, and the appropriate timestep.
Therefore, the optimum values of $N_{pr}$ and $\Delta t_r$ for the IECG simulations also depend on the MD algorithms 
such as the Verlet neighbor list algorithm: in this work, we choose $r_{\rm{skin}}$ values such that
one can use reasonable timesteps with standard frequency of updates of the Verlet neighbor list
for CG simulations as discussed in Section \ref{dtsec}. 
This is because the choice of $r_{\rm{cut}}$ and $r_{\rm{skin}}$ values can depend on the choice of the 
timestep in MD simulations and also on the curvature of the effective CG potential close 
to the $r_{\rm{cut}}$ value.
In the following, we present the results about the effects on the computational efficiency of $N_{pr}$, $\Delta t_r$, and $\alpha$.

\subsection{The Relative Number of Pair Interactions ($N_{pr}$)} 
\label{Nprsec}

\subsubsection{Range and Shape of the Nonbonded Effective CG Potential}
MD simulations involve calculating interactions, and the related forces, between pairs of particles within the range of the potential.
In the IECG model the total effective intermolecular (nonbonded) CG potential, $U^{\rm{inter}}$, is given by
\begin{equation}
U^{\rm{inter}}={\displaystyle \sum_{i}^{n-1}\sum_{j>i}^{n}\sum_\gamma^{n_s}\sum_\alpha^{n_s} U^{ss}_{\rm{eff}} (r_{ij}^{(\gamma\alpha)})},
\label{interucg}
\end{equation}
where $n$, $n_s$, $U^{ss}_{\rm{eff}}$, $r_{ij}^{(\gamma\alpha)}$ are the number of polymers, number of CG sites, 
the effective intermolecular CG potential, and the distance between the CG sites on two different polymers, respectively: 
the Greek indices in eq \ref{interucg} are used to label the number of a CG site along a chain.
The range of the effective CG potential is important because interactions beyond the lengthscales in which
the effective CG potential goes to zero are irrelevant and 
the calculation of the forces can be avoided, thus improving the computational efficiency of the method.

In the simple-minded approach, one defines the radius of a spherical volume, 
the so-called cutoff distance, $r_{\rm{cut}}$,
beyond which forces are not calculated.\cite{Allen1987} 
However, at each MD step, a loop over all pairs of CG particles in the system is required to
calculate the interparticle distances, 
which scales as $\mathcal{O}(n\:n_s)^2$.
The Verlet list algorithm\cite{Verlet1967} further saves CPU time by defining a layer sphere,
called "skin", around the potential cutoff sphere.
While in an initial step a list of neighborhood particles is constructed inside the sphere defined 
by the cutoff distance plus the skin neighbor distance, $r_{\rm{cut}} + r_{\rm{skin}}$, 
in the following few steps distances are only calculated for pairs present in this list.
After a time interval, shorter than the time required for an outside molecule to traverse 
the skin region and get into the range $r_{\rm{cut}}$ of the potential,
the neighbor list is reconstructed and the procedure repeated.

\begin{figure}[htb]
\includegraphics[width=10cm]{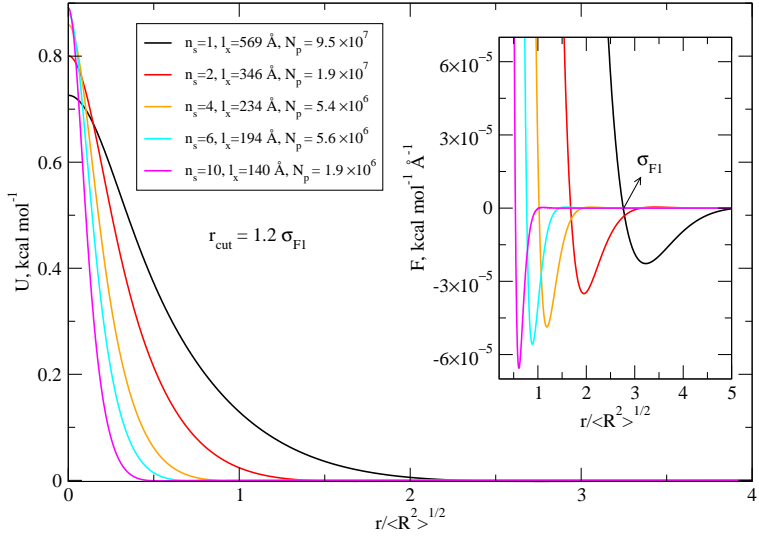}
\caption{The range of potential for a polymer melt with $N=300$ at $503$ K and a monomer density 
of $0.03296$ \AA$^{-3}$ when the polymer is represented by one (black), two (red), four (orange), 
six (cyan), and ten (magenta) CG sites. The inset shows the long-range behavior of the force for all 
CG resolutions and illustrates the first zero of the force $\sigma_{\rm{F1}}$ for the soft sphere model.
The legend shows the values of box lengths ($l_x$) and the number of pairwise interactions ($N_p$)
for various CG resolutions in this study. Note that $r_{\rm{cut}}$s of $1.2 \: \sigma_{\rm{F1}}$ 
were used and the neighbor skin distances are higher for six and ten CG sites models than other 
CG models (see the main text and Table \ref{tabNp} for more details.)
}
\label{ubrange}
\end{figure}

The values of the cutoff and neighbor skin distances at a given state point 
are a function of the range of the intermolecular nonbonded potential, 
which is an important variable in the study of the CG efficiency.
Note that in the CG simulations, the nonbonded potential is a free energy;
therefore, the cutoff distance is state dependent,
which is different from the atomistic simulations, where the cutoff 
distance is not state dependent and usually an intrinsic parameter of the
related force field.
Furthermore, in the CG simulations, as the level of CG increases, the range of the effective CG potential increases, 
which has important effects on the selection of the size of the simulation box.
Finally, the number of pairwise interactions necessary to 
compute forces in the MD simulations is determined by the $r_{\rm{cut}}$ and $r_{\rm{skin}}$ values.\cite{Allen1987} 
A reasonable choice of $r_{\rm{cut}}$ value in the IECG simulations is given by determining the 
first extremum of the effective CG potential (the first zero of the force) over the range of $r>0$, or 
$\sigma_{\rm{F1}}$. Here we select a cutoff distance of $1.2 \: \sigma_{\rm{F1}}$, which allows partial
sampling of the attractive part of the effective CG potential in MD simulations and ensures optimal stability 
of the MD simulations (see below).
An illustration of these effects is presented in Figure \ref{ubrange}, which shows the effective CG potential for a polymer 
with $N=300$ monomers, as the level of CG resolution decreases from a ten CG site representation to the soft sphere model ($n_s=1$).
The inset of Figure \ref{ubrange} shows the effective CG force for the same system while zooming in on the attractive 
component of the effective CG force.

As the level of CG decreases, the effective CG potential at $r=0$ become more repulsive 
and its range decreases such that for the aforementioned polymer melt 
the $r_{\rm{cut}}$ value 
for ten CG site model is about six times less than
the $r_{\rm{cut}}$ value of the soft sphere model, 
but it is still about $3$ to $4$ times larger than
the standard $r_{\rm{cut}}$ values used in atomistic simulations.

In addition, from the analytical solution of the effective CG potential, 
it is known that the effective CG potential is bound 
and its magnitude at contact decreases with increasing the level of CG. 
Furthermore, the range of the potential, normalized by $\sqrt{n_s}$, 
increases with increasing the number of monomers inside the CG unit as $N_s^{1/4}$.\cite{Clark2012}
This scaling behavior can be explained by considering a random walk on the network of 
interpenetrating CG polymers in the space defined by the lengthscale of CG site-CG site interpenetration.
In fact, considering the range $r_{\rm{cut}}$ of the effective CG potential, 
the analytical estimations of its range are in good 
agreement with the numerical ones that are listed in Table \ref{tabNp}.
Our calculations show that the percentage difference range from $1$ to $7$ for 
the soft sphere to ten CG site models. 

\subsubsection{Theoretical Calculation of $N_{pr}$}
It is important to note that coarse-graining has competitive effects on the number of pairwise interactions. 
Considering a liquid of polymers, the number of pairwise interactions between two molecules significantly 
decreases when increasing the extent of CG.
On the other hand, the range of the effective CG  potential increases significantly 
as the level of CG increases, which in turn 
alters the neighbor skin distances and 
increases the cutoff distances
and thus the number of polymers interacting.

Given the cutoff and neighbor skin distances, 
it is possible to theoretically calculate the total number of nonbonded pair interactions used in a given simulation as 
\begin{equation}
N_{p}^{T} = \frac{2 \pi}{3} \rho_s ^2 V (r_{\rm{cut}}+r_{\rm{skin}})^3 , 
\label{Npair}
\end{equation}
where $V$ is the volume of the simulation box, $\rho_s$ is the site density, and $N_{p}^{T}$ is directly related 
to the total number of neighbors in a given MD simulation.
Direct calculations of $N_{p}$ from simulations, which are reported in Table \ref{tabNp} as normalized by the number of sites in the simulation, show that the theoretical predictions are in quantitative agreement with the simulation results, indicating that the number of nonbonded pair interactions are well represented by eq \ref{Npair}.
An analogous quantity is then calculated for the related atomistic (AT) simulation, $N_{p,\rm{AT}}^{T}$,
which is also found to be in quantitative agreement with simulation results.
It is worth mentioning that for the atomistic MD simulations, 
the standard cutoff distance of $14$ \AA\ and neighbor skin distance of $2$ \AA\ are used 
with appropriate system sizes, which have negligible finite size 
effects for the structural and thermodynamical properties such as pressure and 
radial distribution functions (RDFs). 
Therefore, the speed-up in CG simulations due to the relative number of pairwise non-bonded interactions,
which enters eq \ref{Eqeff}, can be directly estimated as $N_{pr}=N_{p,\rm{AT}}^T/N_{p,\rm{CG}}^T$.  
Thus, eq \ref{Npair} is a reliable estimate of the speed-up in efficiency resulting from the change in the
number of pairwise interactions following coarse-graining. 

The legend in Figure \ref{ubrange} shows the number of pairwise 
interactions when cutoff distances of $1.2 \: \sigma_{\rm{F1}}$ (see next section)
and different $r_{\rm{skin}}$ values are used for various CG resolutions,
where the number of pairwise interactions is largest for the soft sphere model, 
which is the highest level of CG for this polymer melt. A more detailed analysis of
the relation between the $r_{\rm{cut}}$, $r_{\rm{skin}}$ and
the number of pairwise interactions in the IECG simulations is presented in Table \ref{tabNp},
which lists the values of $\sigma_{\rm{F1}}$, $\sigma_{\rm{F2}}$, $r_{\rm{cut}}$, $r_{\rm{skin}}$,
suitable system sizes, the number of interaction sites, 
and the number of pairwise interactions in the IECG and atomistic simulations of polymer melts
with degrees of polymerizations of $44$, $192$, and $300$.

\begin{table}[htb]
\centering
\caption{Top:
The number of nonbonded pairwise interactions, $N_{p}$, and the number of interaction sites, $N_{\rm{site}}$,
for several polymer melts with monomer densities of $0.03296$ \AA$^{-3}$ at $503$ K with various degrees of polymerizations ($N$)
in the IECG simulations.
The polymer melts are represented by different number of CG sites, $n_s$, and various system sizes have been used depending
on the appropriate cutoff distances ($r_{\rm{cut}}$) for the IECG simulations (see the main text).
The number of nonbonded pairs per interaction site is shown by $\bar{N_p}$ and
eq \ref{Npair} is used to estimate the theoretical predictions of $N_{p}$ ($N_{p}^{\rm{T}}$).
The box length used in simulations is shown as $l_x$ and $\sigma_{\rm{F1}}$ is the first
zero of the force at a distance range greater than zero, while $\sigma_{\rm{F2}}$ is the second zero.
The $N_{pr}$ is defined as $N_{pr}=N_{p,\rm{AT}}^T/N_{p,\rm{CG}}^T$.
Bottom: The atomistic simulation results are presented for system sizes, which do not exhibit considerable
finite size effects for the properties of polymer melts, such as pressure and pair correlations, at stable state points
in phase diagram.
The statistical uncertainties in the Table are the standard deviations obtained from $10$ independent molecular dynamics simulations.
}
\scalebox{0.8}{
 \begingroup
 \setlength{\tabcolsep}{6pt}
  \begin{tabular}{ccccccccccc}
     \hline
\\
     \multicolumn{11}{c}{ IECG simulations}  \\
     $N$ & $n_s$ & $\sigma_{\rm{F1}}$/\AA & $r_{\rm{cut}}$/$\sigma_{\rm{F1}}$  & $r_{\rm{skin}}$/$\sigma_{\rm{F1}}$ & $l_{x}/\sigma_{\rm{F1}}$ & $N_{\rm{site}}$  & $N_{p,\rm{CG}}^{\rm{T}}$ & $N_{pr}$  & $\bar{N}_{p,\rm{CG}}^{\rm{T}}$ & $\bar{N}_{p,\rm{CG}}^{\rm{sim \: \bullet}}$ \\
     \hline
     $44$ & $1$ & $61$  & $1.2$     & $0.1$      & $2.6$ & $3000$  &  $2.3\times10^6$ & $1.9$ & $781.9$  & $781.9_1$   \\
     $44$ & $1$ & $61$  & $1.85^\ast$    & $0.03$     & $3.8$ & $9261$  &   $2.2\times10^7$ & $0.2$ &  $2384$ & $2393.4_1$  \\
     \hline
     $192$ & $1$ & $156$ & $1.2$    & $0.1$       & $2.6$   & $12000$ & $3.7\times10^7$    & $0.5$ & $2997.3$  & $2998.3_2$   \\
     $192$ & $2$ & $95$  & $1.2$    & $0.1$       & $2.6$   & $5200$  & $7.1\times10^6$    & $2.7$  & $1356.8$  & $1356.8_1$   \\
     $192$ & $4$ & $58$  & $1.2$    & $0.1$       & $2.8$   & $2800$  & $1.7\times10^6$    & $11.2$ & $616.6$  & $614.7_2$   \\
     $192$ & $6$ & $44$  & $1.2$    & $0.1$       & $2.9$   & $2100$  & $8.5\times10^5$    & $22.4$ & $403.5$  & $401.7_2$   \\
     $192$ & $6$ & $44$  & $1.89^\ast$ & $0.05$ & $4.0$ & $5400$ & $7.2\times10^6$  & $2.6$ &  $1330.5$  & $1323.1_3$   \\
     $192$ & $6$ & $44$  & $1.2$    & $0.1$       & $4.0$   & $5400$    & $2.2\times10^6$  & $8.6$ & $405.5$  & $403_1$   \\
     \hline
     $300$ & $1$  & $211$     & $1.2$    & $0.1$        & $2.7$ & $20000$  & $9.5\times10^7$ & $0.3$ & $4745.4$  & $4748.3_2$   \\
     $300$ & $2$  & $128$     & $1.2$    & $0.1$        & $2.7$ & $9000$   & $1.9\times10^7$ & $1.3$ & $2119.2$  & $2119.4_1$ \\
     $300$ & $4$  & $78$      & $1.2$    & $0.1$        & $3.0$ & $5600$   & $5.4\times10^6$ & $4.6$ & $959.5$   & $960.7_1$ \\
     $300$ & $6$  & $59$      & $1.2$    & $0.4$        & $3.3$ & $4800$   & $5.6\times10^6$ & $4.5$ & $1160.8$  & $1159.4_2$ \\
     $300$ & $10$ & $41$      & $1.2$    & $0.4$        & $3.4$ & $3000$  & $1.9\times10^6$  & $13.2$ & $649.2$  & $647.5_2$ \\
     $300$ & $6$  & $59$      & $1.89^\ast$ & $0.02$ & $4.0$ & $8400$  & $1.7\times10^7$     & $1.5$ & $1991.0$  & $1990.1_1$ \\
     $300$ & $6$  & $59$      & $1.2$    & $0.4$        & $4.0$ & $8400$  & $9.8\times10^6$  & $2.6$ & $1160.8$  & $1159.4_1$ \\
     \hline
\\
     \multicolumn{11}{c}{ Atomistic simulations}  \\
     & $N$ & $l_{x}$/\AA  & $r_{\rm{cut}}$/\AA & $r_{\rm{skin}}$/\AA & $N_{\rm{site}}$ & $N_{p,\rm{AT}}^{\rm{T}}$ & $\bar{N}_{p,\rm{AT}}^{\rm{T}}$ & $\bar{N}_{p,\rm{AT}}^{\rm{sim}}$  & &  \\
     \hline
     & $44$  & $78$  & $14.0$  & $2.0$  & $15400$ & $4.4\times10^6$ & $283$  & $279.8_1$ & & \\
     & $192$ & $127$ & $14.0$  & $2.0$  & $67200$ & $1.9\times10^7$ & $283$  & $279.6_1$ & & \\
     & $300$ & $140$ & $14.0$  & $2.0$  & $90000$ & $2.5\times10^7$ & $283$  & $279.6_1$ & & \\
     \hline
  \multicolumn{10}{p{1.0\textwidth}}{
  \textsuperscript{\emph{$\bullet$}}{ The subscripts in the values of $\bar{N}_{p,\rm{CG}}^{\rm{sim}}$ indicate statistical uncertainties in the final digit.
  } } \\
  \multicolumn{10}{p{1.0\textwidth}}{
  \textsuperscript{\emph{$\ast$}}{ The location of the second zero of the force at
  a distance range greater than zero: $\sigma_{\rm{F2}}$.} }  \\
  \end{tabular}
 \endgroup
\label{tabNp}
}
\end{table}
\clearpage

Care is required to estimate $N_{p,\rm{CG}}^{T}$ when one goes to high-resolution CG models 
because as the effective CG potential becomes steeper, 
the force and the particle displacement become larger,
so that relatively larger values of the neighbor skin distance should be used.
Table \ref{tabNp} shows that indeed the $N_{p,\rm{CG}}^{T}$ values for a polymer with 
$N=300$ increases when the CG resolution changes from the four site model to the six site model due to increase in 
the $r_{\rm{skin}}$ value.
However, by performing MD simulations we found empirically that the $r_{\rm{skin}}$ values of
$0.4 \: \sigma_{\rm{F1}}$ work reasonably well for the highest multi-site resolutions of the polymer melt 
with $N=300$ using appropriate timesteps and frequency of updates of Verlet neighbor list as discussed in Section \ref{dtsec};
therefore, the $N_{p,\rm{CG}}^{T}$ values decreases when the CG resolution changes from the six site model to ten site model
as shown in Table \ref{tabNp}. Given the complex nature of the curvature of the effective CG
potential for various levels of CG, we speculate that there is a regime in which relatively
small $r_{\rm{skin}}$ values are not suitable for simulations of high-resolution multi-site CG models
(see the results in Table \ref{tabNp} for the polymer melt with $N=300$).

Furthermore, Table \ref{tabNp} reports the total number of interactions normalized by
the number of sites. The theoretical estimation is given by

\begin{equation}
\bar{N_{p}}^{\rm{T}} =  \frac{2 \pi}{3} \rho_s (r_{\rm{cut}}+r_{\rm{skin}})^3 , 
\label{Npairnorm}
\end{equation}
which does not depend on the volume and/or the simulation boxlength.
Table \ref{tabNp} compares the theoretical estimations with the MD simulation results, 
$\bar{N_{p}}^{\rm{sim}}$, where excellent agreement is observed between
$\bar{N_{p}}^{\rm{T}}$ and $\bar{N_{p}}^{\rm{sim}}$. 
It is important to note that for the atomistic simulations of polymers with various 
degrees of polymerizations at a given monomer density,
the total number of interactions normalized by the number of sites is almost constant  
because the cutoff and neighbor skin distances remain the same. However, one 
can use different appropriate system sizes to simulate the polymer melts, which determine
the efficiency of the atomistic simulations. On the other hand, in the CG simulations
$\bar{N_{p}}^{\rm{sim}}$ varies for various levels of CG because 
the cutoff and neighbor skin distances, as well as the CG site density change, but
the computational efficiency of the CG simulations still depends on the system sizes. 
Therefore, the total number of interactions, $N_{p}$, for the atomistic and CG simulations 
are used to compute  the ratios of the number of pairwise interactions in the atomistic simulations 
to the ones in the CG simulations, $N_{pr}$. 

\begin{figure}[htb]
\includegraphics[width=12cm]{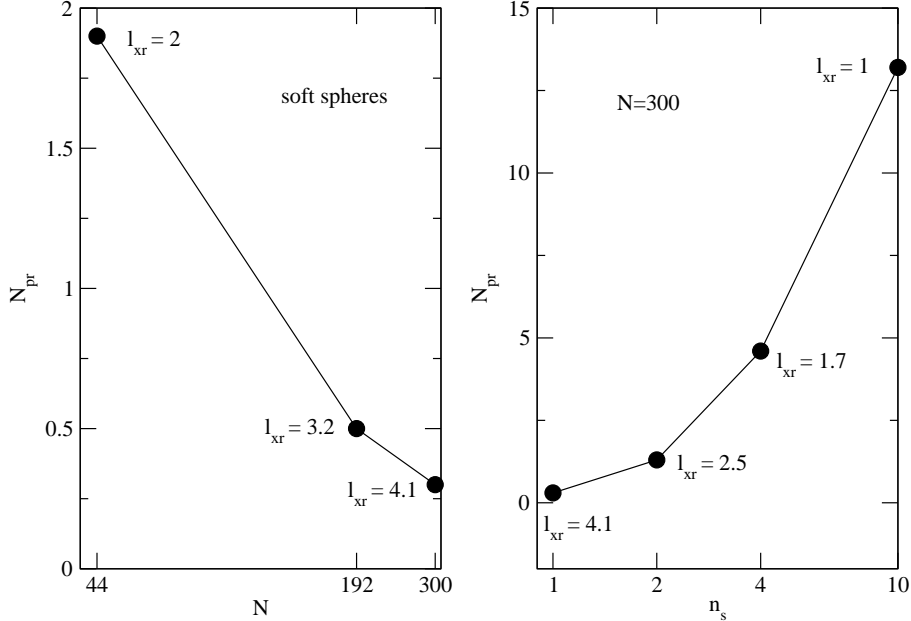}
\caption{The ratios of the number of pairwise interactions in the atomistic simulations to the ones in 
the CG simulations ($N_{pr}$) for various CG resolutions. The ratios of the box lengths in the CG simulations 
to the box lengths in the atomistic simulations ($l_{xr}$) are also presented.
Left panel: The values of $N_{pr}$ as the degree of polymerization ($N$) 
increases when the polymer melts are represented by soft spheres. 
Right panel: The values of $N_{pr}$ as the CG resolution increases
for multi-site models for a polymer melt with $N=300$. 
}
\label{Nprfig}
\end{figure}

Interestingly, the higher the gain in the number of pairwise interactions from atomistic to coarse-graining,
the more convenient the adoption of the CG representation. 
However, the level of CG has competing effects on $N_{pr}$
because increased level of CG implies longer ranged potentials,
and larger simulation boxes, which can lead to a decrease of $N_{pr}$ with increasing level of CG. 
This is illustrated in Figure \ref{Nprfig}, which shows $N_{pr}$ as the degree of polymerization 
increases when the polymers are represented by soft spheres.
Making use of the standard CG system sizes, the left panel of Figure \ref{Nprfig} 
shows that the total number of interactions in the CG simulations
is comparable to the total number of interactions in the atomistic level.
In addition, as discussed before, the range of the effective CG potential for higher CG resolutions is less long-ranged,
which allows using relatively smaller system sizes for the problem of interest. 

The right panel in Figure \ref{Nprfig} shows $N_{pr}$ for $N=300$, which is the highest for ten site model 
although the ten site model has more CG units than the soft sphere models.
This also shows the significance of the range of the effective CG potential in multi-site models,
as also reported in Table \ref{tabNp} where one can see that the range of the effective CG potential is largest
for the soft sphere models of polymer melts. 
Interestingly, this trend becomes opposite at even smaller levels of CG, 
because in the limit of the atomistic resolution (where $n_s=300$) 
it is evident that $N_{pr}=1$ has to be recovered.

Table \ref{tabNp} also reports how the values of $N_{p,\rm{CG}}^{T}$ for various degrees of polymerizations 
and/or CG resolutions increase by an order of magnitude or 
less when the effective CG potential is truncated at the second zero of the force, $\sigma_{\rm{F2}}$,
which can significantly decrease the computational efficiency.
Therefore, it is important to address what values 
of $r_{\rm{cut}}$ are suitable to enhance the computational efficiency, 
which is addressed in more detail in the next section.

\subsubsection{Optimization of the Cutoff and Neighbor Skin Distances}

Using standard MD simulation parameters, it is observed that truncating the effective CG potential at 
$\sigma_{\rm{F1}}$ at various state points can introduce significant artifacts especially for the 
multi-site models (results not shown); therefore, 
to report the efficiency while making sure the MD simulations results are accurate, 
the effective CG potential is truncated at greater distances than $\sigma_{\rm{F1}}$,
($1.2 \: \sigma_{\rm{F1}}$) to allow exploration of the attractive part of the effective CG potential.

\begin{figure}[htb]
\includegraphics[width=10cm]{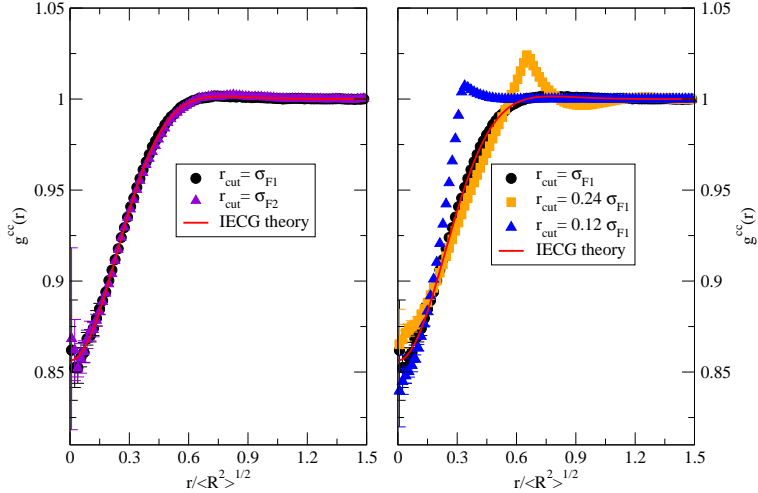}
\caption{
Illustration of various truncations of the effective IECG potential as it appears in the 
radial distribution function of soft spheres representing a polymer melt with $N=300$
at $503$ K and the monomer density of $0.03296$ \AA$^{-3}$.
The cutoff distances are in units of the first zero of the force, $\sigma_{\rm{F1}}$.
See the legends for the values of the cutoff distances, $r_{\rm{cut}}$.
}
\label{rcutRDF}
\end{figure}

The left panel of Figure \ref{rcutRDF} shows consistent RDFs 
for the polymer melts with $N=300$ represented by soft spheres at $503$ K at a monomer density of
$0.03296$ \AA$^{-3}$ when the IECG effective potential
is truncated at $\sigma_{\rm{F1}}$ and $\sigma_{\rm{F2}}$, both of which are consistent with the
IECG theory predictions. It is worth mentioning that the RDFs calculated for the given samples 
are consistent with the atomistic simulations.\cite{Dinpajooh2017}
The right panel of Figure \ref{rcutRDF} shows that the inaccurate truncation of the IECG effective potential 
can result in significant errors in RDFs. The blue triangles show the RDF when the effective CG potential
is truncated at a distance of about $0.12 \: \sigma_{\rm{F1}}$, which results in
significant deviations in the correlation holes as compared to the accurate RDF.
The deviations in the correlation hole is less pronounced when the effective CG potential is truncated
at a distance of about $0.24 \: \sigma_{\rm{F1}}$ (orange squares), 
but artificial peak and minimum appear in the RDF.
Making use of a distance of about $0.5 \: \sigma_{\rm{F1}}$ to truncate the 
potential still introduces an artificial peak but less pronounced than
the red squares (results not shown). 

Note that for the homogeneous systems studied in this work, 
the long-ranged CG corrections\cite{Dinpajooh2017} can be used 
to accurately compute the properties of interest such as pressure.
In fact, the tail CG corrections turn out to work reasonably well for the estimation of the pressure 
obtained from various truncation of the effective CG potential when the MD simulations 
are stable.
For instance, the values of pressure corrected with the CG long-range contributions
for the relevant data presented in Figure \ref{rcutRDF} are about $343$ atm,
which are in agreement with the atomistic simulation results (See Figure \ref{RDFTcons}).
This advancement in treating the long-range CG interactions in the IECG simulations reasonably results in the 
consistencies of various levels of CG with the atomistic simulation results in properties such as 
RDF and pressure. 

In addition, Table \ref{tabNp} shows that larger values of $r_{\rm{skin}}$ are used for 
the IECG simulations of high-resolution multi-site CG models than for the soft sphere model, for a polymer melt with 
$N=300$. This is because as the level of CG decreases and the range of the effective CG potential decreases, 
the curvature of the repulsive part of the effective CG potential becomes sharper such that relatively
larger values of $r_{\rm{skin}}$ are required to avoid the failure of the MD algorithm. 
As discussed in Section \ref{rhotemp}, for a polymer melt represented by a given CG resolution,
the curvature of the effective CG potential alters significantly less with varying density and temperature.
This suggests that the optimum $r_{\rm{skin}}$ values do not change considerably for a polymer melt at a given degree 
of polymerization, in the range of temperature and density studied here.

Note that the system sizes in the standard IECG simulations is determined from 
the appropriate $r_{\rm{cut}}$ and $r_{\rm{skin}}$. 
If one is not interested in problems that require large-scale simulations as discussed above, 
the boxlength in the IECG simulation can be set to be somewhat greater than twice 
as much as $r_{\rm{cut}}$, which suggests that the number of pairwise interactions
would be highest for the soft sphere models. 
In Table \ref{tabNp} the values of the box length are reported as $l_{x}$.
As the CG resolution increases, the range of potential decreases and at a given CG resolution
one may use the system sizes close to the system sizes in the atomistic simulations such as the
ten CG site model of polymer melt with $N=300$ as shown in Table \ref{tabNp} and Figure \ref{Nprfig}.

It is worth noting that a relatively small value of $r_{\rm{skin}}$ is required when 
the effective CG potential is truncated at larger distances, 
such as at  the second zero of the force $\sigma_{\rm{F2}}$.
This is mainly 
due to the large volume accessible to store the Verlet neighbor list when the 
effective CG potential is truncated at large distances and also the 
change in the curvature of the effective CG potential close to $\sigma_{\rm{F2}}$.

To end this section, it is worth mentioning that as a consequence of applying the Verlet neighbor list, 
combined with the link-cell method, the CPU time for computation of forces per MD timestep in the IECG simulations scales linearly 
with the number of IECG sites.\cite{Plimpton1995,Chialvo1990,Meyer2014}

\subsubsection{Density and Temperature Dependence: Range of the Effective CG Potential}
\label{rhotemp}

\begin{figure}[htb]
\includegraphics[width=12cm]{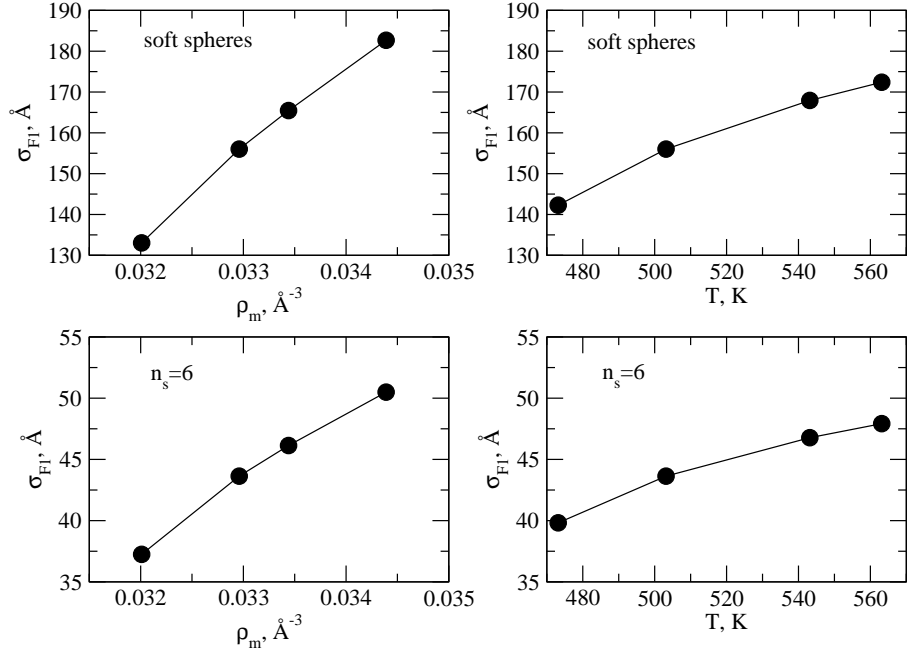}
\caption{ Values of the first zero of the force over the distances greater than zero, $\sigma_{\rm{F1}}$,
for polymer melts with a degree of polymerization of $192$ at different densities and temperatures.
Left panels: Change in $\sigma_{\rm{F1}}$ for polymer melts at $503$ K as the density changes when
it is represented by soft sphere or six CG site models.
Right panels: Change in $\sigma_{\rm{F1}}$ for polymer melts at a monomer density of $0.03296$ \AA$^{-3}$
as the temperature changes when it is represented by soft sphere or six CG site models.
}
\label{Trhodep}
\end{figure}

Noting that the effective IECG potential is in fact a free energy,
as the density and temperature increases for a given polymer melt,
the first zero of the force, $\sigma_{\rm{F1}}$, increases in the IECG approach
(see our previous work for the details\cite{Dinpajooh2017}).
Figure \ref{Trhodep} illustrates this for a polymer melt with $N=192$, when it is
represented by soft sphere and six CG site models.
This has implications on the computational efficiency of the IECG simulations.
Making use of eq \ref{Npair} and
estimating the standard volume of the simulation box as $(2.5 \: \sigma_{\rm{F1}})^3$ and
the $r_{\rm{cut}}$ and $r_{\rm{skin}}$ values $1.2 \: \sigma_{\rm{F1}}$ and $0.1 \: \sigma_{\rm{F1}}$,
respectively, the number of nonbonded pairwise interactions can be calculated as follows:

\begin{equation}
N_{p}^{T} = A \rho_s ^2 \sigma_{\rm{F1}}^6 , 
\label{Nsigma}
\end{equation}
where $A \approx 80$, and $\rho_s$ is the site density.

Therefore, at a given temperature as the site density
and consequently $\sigma_{\rm{F1}}$ increases, the number of pairwise
interactions in an IECG simulation can increase.
This increase is more pronounced for the soft sphere models than the
six CG site models. Similarly, at a given density, as the temperature
increases so does $\sigma_{\rm{F1}}$; therefore, the
number of pairwise interactions in an IECG simulation can increase.
It is worth noting that the curvature of the IECG effective
potential slightly changes at various temperatures and densities, which
allows one to use the close values to the optimum neighbor skin distances reported in
Table \ref{tabNp}.

\subsection{Timestep ($\Delta t$)}
\label{dtsec}

The equation of motions in MD simulations are integrated using finite difference algorithms, such
as the Verlet integration algorithm\cite{Allen1987}, 
which can lead to errors if the suitable timestep is not chosen, 
because making use of very large timesteps can cause numerical instabilities. 
Meanwhile, the larger the timestep, the more rapid sampling of the phase space,
and the longer the timescale reached by the simulation.
The MD timestep may be selected to be one or two orders of magnitude smaller than the characteristic
timescale,$\tau$, of the fastest motion of the system. 
In atomistic simulations, it is typical to adopt a timestep of femtoseconds, as will be discussed below.
Similarly, the timestep in a given CG-MD simulation should be smaller than the fastest motion in the CG system. 

Different CG models afford different timesteps, depending on the level of resolution of the CG model.
We compare the IECG models in soft sphere and multi-site representations to other models of polymers,
including the united atom models,\cite{Paul1995,Ramos2015}
the MARTINI CG models,\cite{Marrink2007} and the CG model by Grest and coworkers.\cite{Salerno2016}
Starting from all atom MD simulations of simple polymeric systems, 
where the shortest lengthscale is the atomistic lengthscale of $0.1$ nm, 
the fastest motions of the system involve C-H vibrations,
which restricts one to use timesteps of $0.5-1$ fs.\cite{Muller-Plathe2002,Siu2012}
In the united atom models, the fast vibrations of the C-H bonds are averaged out, 
and the CH, CH$_2$ and CH$_3$ atomic groups are treated as single rigid particles, allowing one
to have access to about $1$ nm lengthscale resolutions and to use a timestep of $1-2$ fs
in standard MD simulations.\cite{Ramos2015}
It is worth mentioning that multiple timestep methods such as
Reversible REference System Propagator Algorithm (RESPA)\cite{Tuckerman1992} 
can yield $2-3$ times speed-up in MD of all atom or united atom simulations, 
but they are less efficient in parallel MD simulations due to CPU communication.
Neglecting the hydrogen atoms and on average representing
four heavy atoms by a single interaction site,
the MARTINI force field,\cite{Marrink2004,Marrink2007,Murtola2009,Marrink2010a,Marrink2010} developed 
to study biomolecular systems, allows for timesteps of $20-30$ fs 
in MD simulations, when an appropriate update of the Verlet neighbor
lists is adopted. Recently, Grest et al. developed a CG model,
which groups from two to up to six CH$_2$ units into one CG unit,
giving access to timesteps ranging from $2$ to $10$ fs
in their MD simulations.\cite{Salerno2016} 
Compared to the IECG models, the aforementioned CG models are low-level CG models.  

\begin{figure}[htb]
\includegraphics[width=12cm]{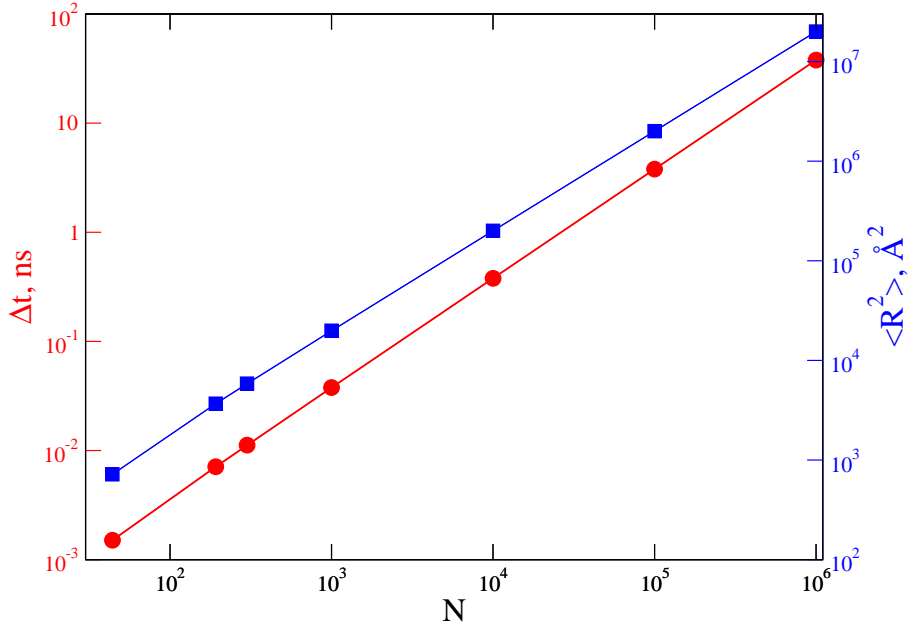}
\caption{Values of timestep, $\Delta t$, (red circles) and $\langle R^2 \rangle$ (blue squares) computed for polymer 
melts of various degrees of polymerization, $N$, at $503$ K and a monomer density of $0.03296$ \AA$^{-3}$ for the soft sphere IECG model.} 
\label{R2dt}
\end{figure}

In the IECG simulations, the appropriate timesteps may be estimated from 
the characteristic timescales as $\Delta t \approx 0.05 \tau$.
For the soft sphere models, the characteristic timescale is given as 
\begin{equation}
\tau_{\rm{soft}} = \left( \frac{M \langle R^2 \rangle }{k_{\rm{B}} T} \right)^{1/2},
\label{tausoft}
\end{equation}
where $M$ is the mass of soft sphere and $\langle R^2 \rangle$ is the mean-square end-to-end polymer distance. This time is related to
the mean time between soft sphere collisions. 
Depending on the degree of polymerization ($N$) and the level of molecular CG, 
this characteristic timescale for soft sphere models may range from picosecond for $N \approx 30$ to microsecond
for $N \approx 10^6$; therefore, as shown in Figure \ref{R2dt}, appropriate theoretical timesteps 
may range from hundreds of femtoseconds to tens of nanoseconds in the IECG simulations.
This is another advantage of the IECG method in comparison with the high-resolution CG methods discussed above, 
where the timestep usually does not exceed tens of femtosecond. 

In the practical application of MD simulations, one has to carefully consider
the Verlet neighbor list updates to set up the appropriate timesteps, which 
should obey the following relationship:

\begin{equation}
(n_t - 1 ) \Delta t \:  v \leq r_{\rm{skin}},
\label{verleteq}
\end{equation}
where $ n_t \geq 1$ is the number of timesteps between two consecutive updates,
$v$ is the velocity of a given CG site, which directly depends on the temperature. 

Considering the frequency of updating the Verlet neighbor lists, CG timesteps larger than
the characteristic timescale will result in the traveling of soft particles
through distances larger than the (sub)domain, in a time prior to the (sub)domain calculation, which
distorts the trajectory in phase space of the system, and affects the properties calculated 
from time averages. 
In other words, when the Verlet lists and the subdomains are built, 
for instance in the combined Verlet list and link-cell algorithms, 
particles may be lost if a large timestep is used and a particle travels a long distance before the list is updated.
Therefore, the suitable timestep also depends on the sizes of the 
subdomain, which is related to the $r_{\rm{skin}}$ and $r_{\rm{cut}}$ values
(see eq \ref{verleteq}).

In addition, truncating the potential at short $r$ leads to a more repulsive
potential, so that
in the given timestep the high repulsive forces can make the particles travel unphysically long 
distances, with consequent failure of the MD algorithm.  Therefore, in similar conditions, as the $r_{\rm{cut}}$ value increases 
and the contribution of the attractive part of the potential is properly taken into account in the IECG method, 
a larger CG timestep can be used with less likelihood of the failure of the MD algorithms.

Our MD simulations suggest that an $r_{\rm{cut}}$ value 
of $1.2 \: \sigma_{\rm{F1}}$ allows one to use the appropriate CG timesteps 
($\Delta t \approx 0.05 \tau$) using the standard update of the Verlet neighbor list in the IECG simulations.
To assess the accuracy of the suitable timestep, a series of IECG-MD simulations were performed 
starting from the same equilibrated liquid configurations using the Nos$\rm{\acute{e}}$-Hoover thermostat. 

\begin{figure}[htb]
\includegraphics[width=12cm]{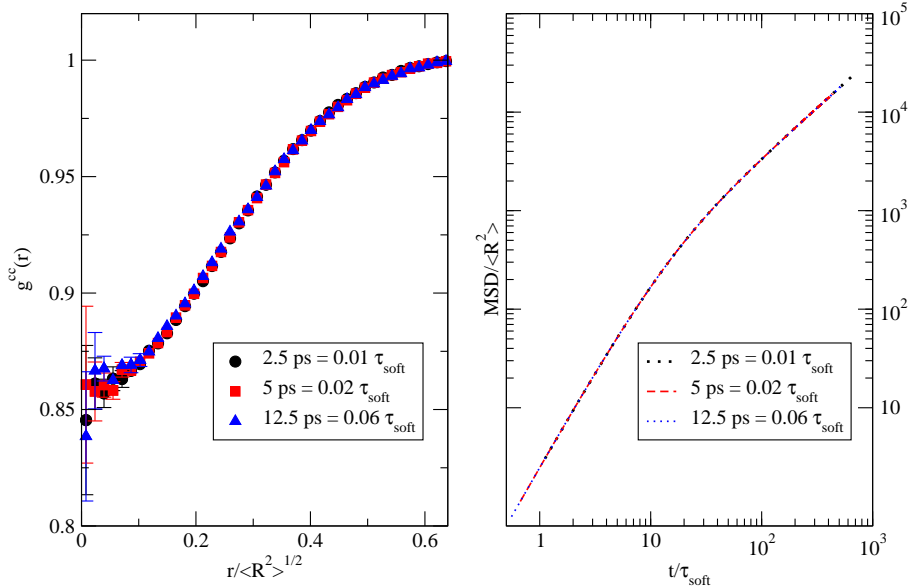}
\caption{Left Panel: Radial distribution function of soft spheres representing a polymer melt with $N=300$
at $503$ K and the monomer density of $0.03296$ \AA$^{-3}$.
Timesteps ($\Delta t$) of $2.5$ ps (circles), $5$ ps (squares), 
and $12.5$ (triangles) ps have been used in molecular dynamics simulations.
For comparison the timesteps are presented in terms of $\tau_{\rm{soft}} \approx 224$ ps. 
Right Panel: The mean square deviation of the aforementioned soft sphere models. 
}
\label{dtrdf}
\end{figure}

Pressure, RDFs, and mean-square displacements were tested. Figure \ref{dtrdf} shows consistencies in the RDFs and 
mean square displacements when relatively small and large 
timesteps have been used 
in the IECG simulations of polymer melts with degrees of polymerization of $300$ at $503$ K and 
the monomer density of $0.03296$ \AA$^{-3}$ with the polymer represented as one soft sphere. 
In addition, the IECG simulations result in the pressure value of $343.3$ atm,
kinetic energy per particle of $1.5$ kcal mol$^{-1}$,
potential energy per particle of $44.4$ kcal mol$^{-1}$, and temperature of
$503$ K when timesteps less than $12.5$ ps and reasonable values of $r_{\rm{skin}}$ are used.   
Therefore, the IECG simulations show that making use of 
larger timesteps up to $12.5$ ps is possible for a polymer melt with $N=300$ when represented by 
soft spheres, with the caveat that one has to use reasonable values of $r_{\rm{skin}}$ 
and Nos$\rm{\acute{e}}$-Hoover thermostat parameters. 
Note that other common thermostats such as Berendsen thermostat may also be used in the IECG simulations.
It is worth mentioning that in cases where common thermostats are not efficient, 
a simple velocity rescaling method can allow one to get reasonable results (not shown here) 
while saving computational time due to the simplicity of the algorithm.
This is because all ensembles in the thermodynamic limit should give consistent results.

In the multi-site CG model, the characteristic timescale corresponds to the highest 
frequency motion of the CG bond vibrations, where the effective bond potential in the IECG method 
for a given polymer melt, $i$, is then given by 

\begin{equation}
U^{\rm{bond}}_i = \sum_\gamma^{n_s-1} \left( 2 n_s k_{\rm{B}}T l_{i\gamma}^2/< R^2 > + k_{\rm{B}} T U^{ss}_{\rm{c}}(l_{i\gamma}) \right) \ , 
\label{ubond}
\end{equation}
with $n_s$ the number of CG sites in which the polymer chain is partitioned,  $l$ the bond length
between CG sites, and $U^{ss}_{\rm{c}}$ the correction term that enforces the correct distributions
between adjacent CG sites.

Approximating the force constant of the multi-site bond vibrations by the first term in eq \ref{ubond},
i.e., treating the effective potential as a harmonic oscillator, gives a relaxation time for the fastest modes of vibration, $\tau_{\rm{site}}$, which theoretically scales as

\begin{equation}
\tau_{\rm{site}} \propto \frac{ \tau_{\rm{soft}} }{n_s} \ .
\label{tausite}
\end{equation}

Note that in the IECG simulations, the bonded sites are given a bond potential derived from the
direct Boltzmann inversion of the probability distribution of the effective
bond length, where the details are discussed in our previous works.\cite{Clark2013,McCarty2014}
Therefore, treating the effective potential as the harmonic oscillator, the relaxation times of the fastest modes of vibration,
$\tau_{\rm{site}}$, theoretically scales as $1/n_s$.
However, in practice the choice of appropriate timestep also involves the range of the effective CG potential, the cutoff distances,
the neighbor skin distances, and the frequency of updating the Verlet list. Therefore, it is useful to evaluate if the theoretical scaling is recovered directly from MD simulations.

\begin{figure}[htb]
\includegraphics[width=12cm]{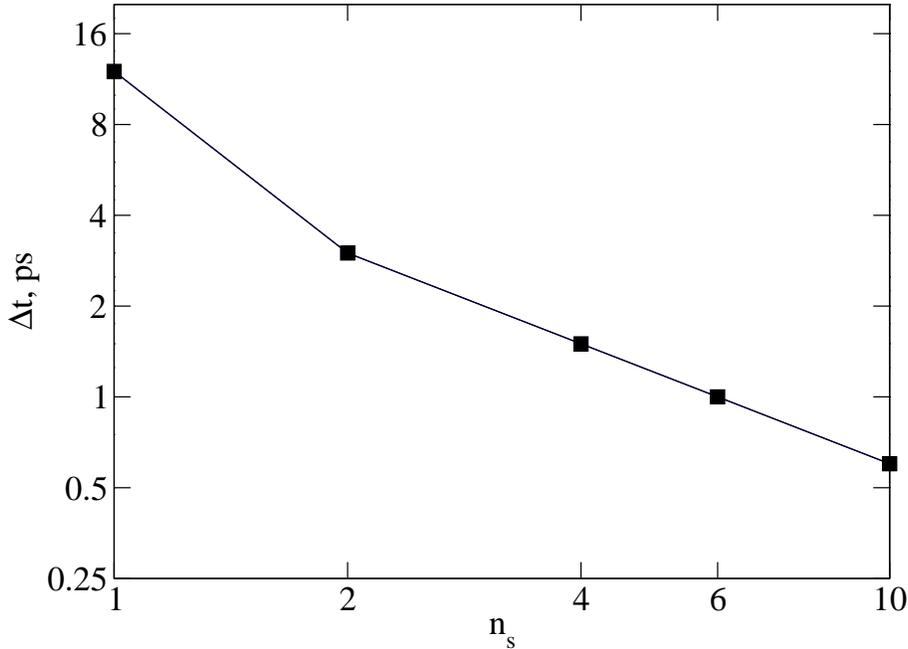}
\caption{
The effect of CG resolutions on the practical timestep is presented for multi-site models
for polymer melts with degree of polymerization $300$ at $503$ K with a monomer density of $0.03296$ \AA$^{-3}$.
}
\label{nbdt}
\end{figure}

The effect of CG resolutions on the timestep was empirically investigated for multi-site models 
in the IECG simulations using the update frequency mentioned in Section \ref{simsec}.
Making use of the combined Verlet neighbor list and link-cell algorithms\cite{Plimpton1995}, the maximum CG timesteps that don't lead to numerical instability are observed in MD simulations for various levels of CG, while using one processor. Figure \ref{nbdt} shows the effect of
the CG resolutions on this empirical CG timestep for the multi-site models of polymer melts with $N=300$ at a given state point.
Making use of the standard cutoff distances and neighbor skin distances presented in Table \ref{tabNp}, the empirical timestep is found to obey the theoretical scaling with the number of CG sites (eq. \ref{tausite}).

\subsection{Dynamical Scaling Factor ($\alpha$)}
\label{alphasec}

In principle, the CG procedure results in relatively much smoother free energy surfaces 
than the atomistic description because local degrees of freedom are averaged out during coarse-graining.
In addition, coarse-graining on various levels results in various shapes of CG units with
different hydrodynamic radii and friction coefficients.\cite{Lyubimov2010,Lyubimov2011,Lyubimov2013} 
Both contributions speed up the dynamics, which significantly enhance sampling of the phase
space for a given system.

% $$$$$$$ 
Starting from the first-principle Liouville equation  and applying the Mori-Zwanzig projection 
operator technique,\cite{Zwanzig2001}  we derived Generalized Langevin Equations (GLE) for the atomistic and for the IECG representation.\cite{Lyubimov2010,Lyubimov2011,Lyubimov2013}
The two GLEs differ in the linear term, which contains the entropic contribution due to the smoothing of the free energy landscape following CG,
and in the memory function, which leads to different site friction coefficients. 
Both corrections are essential to reconstruct the atomistic dynamics from the IECG-MD simulation. 
%In future work, the method will be applied to multisite models. 

When accurate GLE formalisms are not formulated for the atomistic and CG representations,
it is common practice to calculate empirically a numerical scaling factor by
directly comparing dynamical quantities such as the mean square displacement (MSD), 
or diffusion coefficients, in the CG and
atomistic simulations.\cite{Fritz2009,Fritz2011,Salerno2016} 
Note that this empirical procedure doesn't properly account for the physical motivation of the different dynamics that appear 
at varying levels of CG. However, given that we are interested only in determining the numerical gain 
following coarse-graining, we adopt this simple procedure to calculate the speed-up factors.

\begin{figure}[htb]
\includegraphics[width=1\columnwidth]{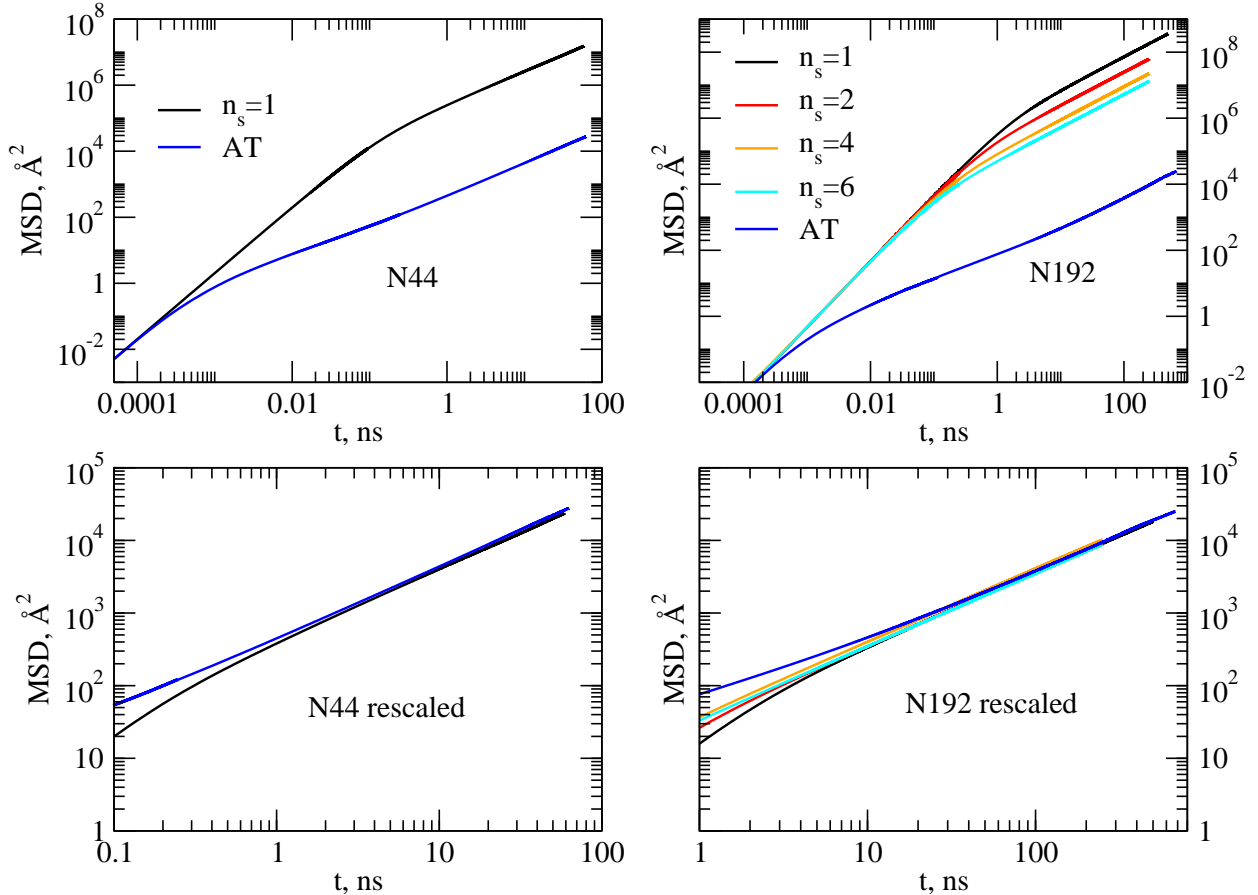}
\caption{The top left and right panels show the center of mass mean square displacements 
(MSD) for the IECG simulations at various CG resolutions and the atomistic simulations of
the polymer melts with degrees of polymerizations of $44$ and $192$, respectively.
The bottom panels display the rescaled MSD of the IECG simulations with respect to the atomistic simulations. 
}
\label{MSDrescale}
\end{figure}

This requires running relatively long atomistic trajectories of polymer melts to reach the diffusive
regime. Figure \ref{MSDrescale} demonstrates this for polymer melts with degrees of polymerizations 
of $44$ and $192$ in the relatively long atomistic simulations
and the IECG simulations at various levels of CG. 
The diffusion coefficient, $D$, in a system of identical particles can be computed 
from the linear part of the center of mass MSD as a function 
of time, $t$, using the Einstein relation:\cite{Hansen2003}

\begin{equation}
D = \lim_{t \rightarrow \infty} \frac{\langle \left( R_{\rm{cm}}(t) - R_{\rm{cm}}(0)  \right)^2  \rangle}{6 t}, 
\label{diff}
\end{equation}
where $R_{\rm{cm}}$ is the position of the center of mass of a given polymer.
The dynamical scaling factor can then be estimated as the ratio between the diffusion 
coefficients from the CG and atomistic (AT) simulations, i.e. $\alpha=D_{\rm{CG}}/D_{\rm{AT}}$.
We use this approach to compute the dynamical
scaling factor for relatively short chain polymer melts. 
The results are presented in Table \ref{tabeff}, where the 
diffusion constants in atomistic simulations were calculated as 
$74.1$, $6.3$, and $2.4$ \AA$^2$/ns for the polymer melts 
with degrees of polymerizations of $44$, $192$, and $300$, respectively.

\subsection{Overall Computational Efficiency}
\label{effsec}

\begin{table}[tbh]
\centering
\caption{The efficiency of CG simulations compared to the atomistic ones, when eq \ref{Eqeff} is used.
The values of $\Delta t_r$ are the ratios of the CG timestep to the atomistic timestep.
The values of $N_{pr}$ are the ratios of the CG non-bonded pairwise forces to the atomistic ones, 
when cutoff distances of $1.2 \: \sigma_{\rm{F1}}$ and the neighbor skin distances reported in Table \ref{tabNp} 
are used. Nos$\rm{\acute{e}}$-Hoover thermostat is used for all molecular dynamics simulations.} 
\begin{tabular}{cccccc}
     \hline
     $N$   & $n_s$ & $\alpha$ & $\Delta t_r$ & $N_{pr}$ & $\varepsilon$ \\
     \hline
     $44$  & $1$   & $6.5\times10^2$  & $900$   & $1.9$ & $1.0\times10^6$ \\
     \hline
     $192$ & $1$   & $2\times10^4$    & $3500$ & $0.5$  & $3.3\times10^7$ \\
     $192$ & $2$   & $7\times10^3$    & $800$  & $2.7$  & $1.4\times10^7$ \\
     $192$ & $4$   & $2\times10^3$    & $400$  & $11.2$ & $1.1\times10^6$ \\
     $192$ & $6$   & $1.5\times10^3$  & $250$  & $22.4$ & $7.8\times10^6$ \\
     \hline
     $300$ & $1$   & $7\times10^4$    & $6000$ & $0.3$  & $1.1\times10^8$ \\
     $300$ & $2$   & $2\times10^4$    & $1500$ & $1.3$  & $4.0\times10^7$ \\
     $300$ & $4$   & $8\times10^3$    & $750$  & $4.6$  & $2.5\times10^7$ \\
     $300$ & $6$   & $3.5\times10^3$  & $500$  & $4.5$  & $9.7\times10^6$ \\
     $300$ & $10$  & $2\times10^3$    & $300$  & $13.2$ & $9.1\times10^6$ \\
     \hline
\end{tabular}
\label{tabeff}
\end{table}

Overall, the computational efficiency of the IECG simulations,
$\varepsilon$, can be reported as defined in eq \ref{Eqeff}, and reported in Table \ref{tabeff}.
The dynamical scaling factor turns out to dominate the other terms in eq \ref{Eqeff} 
while considerable computational efficiency can be gained via the enhancement of the timestep
in the IECG simulation. The computational gain based on the number of pairwise interactions, 
$N_{pr}$, can have positive and negative effects based on the level of CG or the CG resolution
noting that the contribution of this term among other terms is lowest in eq \ref{Eqeff}.
The range of potential significantly increases for soft sphere 
models such that the computational gain based on $N_{pr}$
is inverse for the polymer melts with $N=192$ and $300$ when represented by soft spheres, but 
the computational gain of $1.9$ is achieved for the soft sphere model of polymer melt with $N=44$.

For soft sphere models, as the level of CG increases, the dynamical scaling factor and the 
CG timesteps increase while the computational gain via $N_{pr}$ in most cases decreases due to the long-range 
CG potential. For a given polymer melt, as the CG resolution increases
the CG timesteps decrease, but the computational gain via $N_{pr}$ requires care because 
it depends on the change in the $r_{\rm{cut}}$ values as well as in the $r_{\rm{skin}}$ values.
In general, as the CG resolution increases, the $r_{\rm{cut}}$ values decrease, but the $r_{\rm{skin}}$ values,
which are directly related to the optimum values of the CG timestep, can increase and compensate
the computational gain obtained from the decrease in the $r_{\rm{cut}}$ values. 

Although $N_{pr}$ for the polymer 
melt with $N=192$ decreases as the CG resolution decreases, $N_{pr}$ has a more complex behavior 
for the polymer melt with $N=300$ such that the $N_{pr}$ values for four and six CG site models 
are almost equal due to the significant change in the $r_{\rm{skin}}$ value for the six CG site model. 
At the highest level of granularity, the CG model needs to recover the atomistic behavior.
In a nutshell, Table \ref{tabeff} shows that the overall computational efficiency is highest for the 
soft sphere models and lowest for the highest resolution multi-site models. 

\begin{figure}[htb]
\includegraphics[width=14cm]{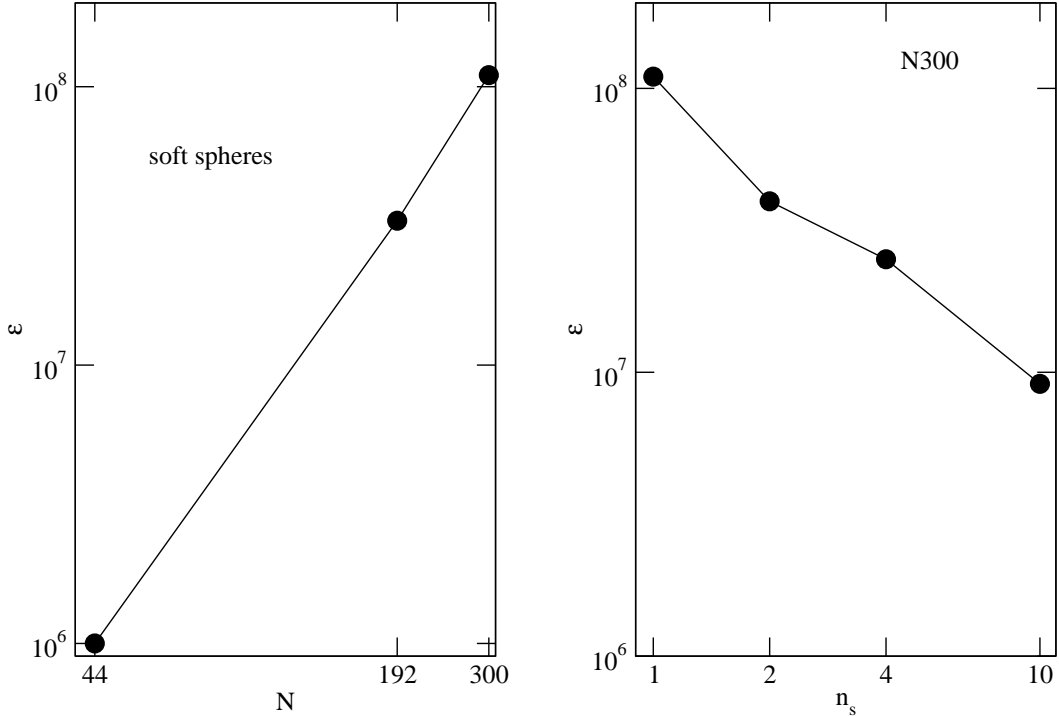}
\caption{Efficiency values, $\varepsilon$, as reported by eq \ref{Eqeff} for various levels 
of CG.  
Left panel: Efficiency values as the degree of polymerization ($N$) increases when the polymer melts are represented by soft spheres. 
Right panel: Efficiency values for a polymer melt with $N=300$ 
as the CG resolution changes from soft sphere model to ten CG site model. 
}
\label{toteff}
\end{figure}

Figure \ref{toteff} summarizes the overall computational efficiency results. The left panel of 
Figure \ref{toteff} shows millions magnitude computational efficiency for the soft
spheres as the degree of polymerization increases while the right panel shows that 
the overall efficiency for a polymer melt with $N=300$ at various CG resolutions, which 
decreases as the level of CG decreases.  

\subsubsection{Parallel Scalability}
\label{pareff}

Due to the importance of parallel computing,
we complete our study with a discussion on the parallel scalability of the IECG simulations
as compared with the atomistic simulations. 
A thorough description of fast parallel algorithms such as
atom decomposition, force decomposition, spatial decomposition for MD simulations are given 
in ref \citen{Plimpton1995}. All three aforementioned methods balance computation optimally
as $N_{\rm{site}}/(2P)$, where $N_{\rm{site}}$ is the number of interaction sites 
and $P$ is the number of processors and Newton's third law is implemented 
for interactions between particle pairs inside a processor's box. 
 
Standard spatial decomposition algorithms subdivide the physical simulation domain into
small three-dimensional boxes for each processor and uses two data structures: 
one for storing the $N_{\rm{site}}/P$ particles in its box, which contains a complete set of data
such as position, velocity, and Verlet neighbor lists, and the other one for particles
in nearby boxes, which only stores particle positions and the six data exchange algorithm described 
in ref \citen{Plimpton1995}, which is used for exchanging all particle positions in adjacent boxes.  

Added to the computation, which scales as $N_{\rm{site}}/(2P)$, 
the communication scales as $(N_{\rm{site}}/P)^{2/3}$ for relatively large problems
and the memory scales as $N_{\rm{site}}/P$. 
Unlike the link-cell algorithm, 
the box length (sub-domain) may now be smaller, equal, or larger than the $r_{\rm{cut}}$ value.
This has important implications for the IECG simulations 
because the long-range nature of the IECG effective potential requires using
large $r_{\rm{cut}}$ values. Therefore, the sub-domain of a processor can be 
smaller than the $r_{\rm{cut}}$ value and the particle information is required
from more distant boxes, which requires sending $N_{\rm{site}}/P$ positions of a sub-domain 
to many neighbors and extra communication to acquire the required particle positions and forces.

Note that each processor computes forces on particles in its sub-domain 
using the ghost information from nearby processors, and processors communicate and store ghost particle information
for particles that border their sub-domain.
On the other hand, as discussed in section \ref{Nprsec}, as the level of CG decreases, the range of the effective 
CG potential decreases, and the total number of interactions can decrease in the IECG simulations. Therefore, 
it is more likely to lead to the communication overhead, which can lower/stop the speed-up for a given number of CPUs. 

\begin{figure}[htb]
\includegraphics[width=14cm]{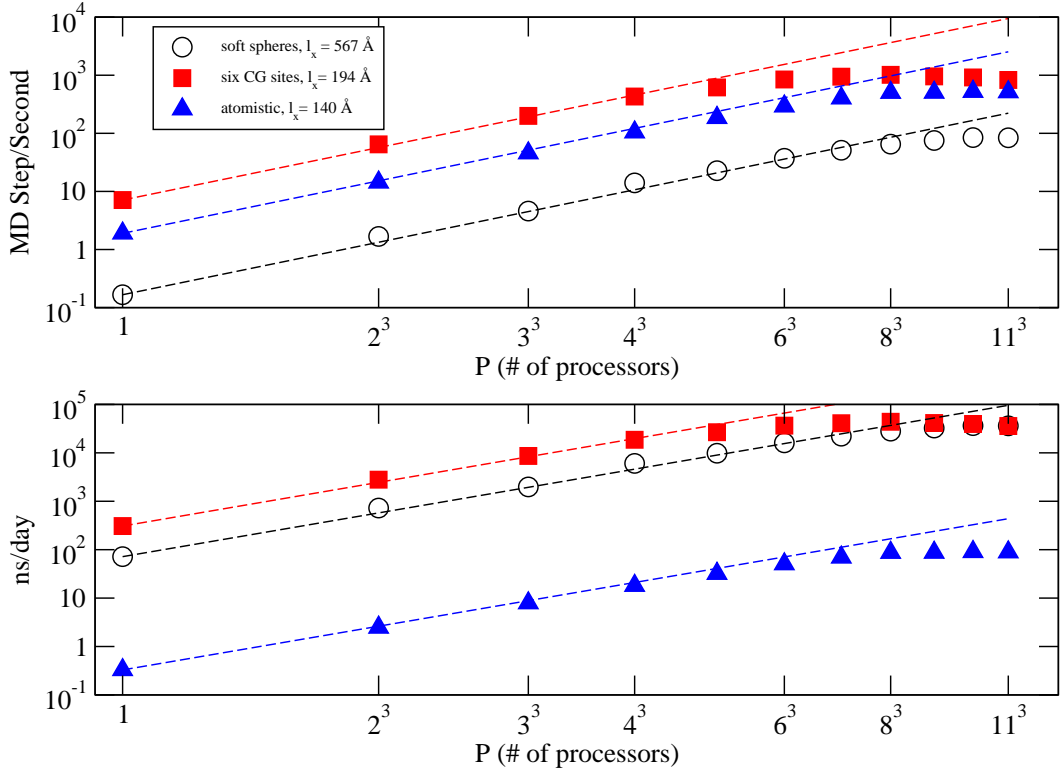}
\caption{Strong scaling performance for atomistic (blue triangles), soft sphere (black circles), 
and six CG site (red squares) models of polymers with degrees of polymerization of $300$ at $503$ K
and a monomer density of $0.03296$ \AA$^{-3}$:
Top panel: The performance is reported in terms of molecular dynamics step per second. 
Bottom panel: Including the timestep in the performance, the performance is 
reported in terms of nanosecond per day.
Note that the statistical uncertainties are smaller than the symbol sizes. 
}
\label{scalability}
\end{figure}

\begin{figure}[htb]
\includegraphics[width=14cm]{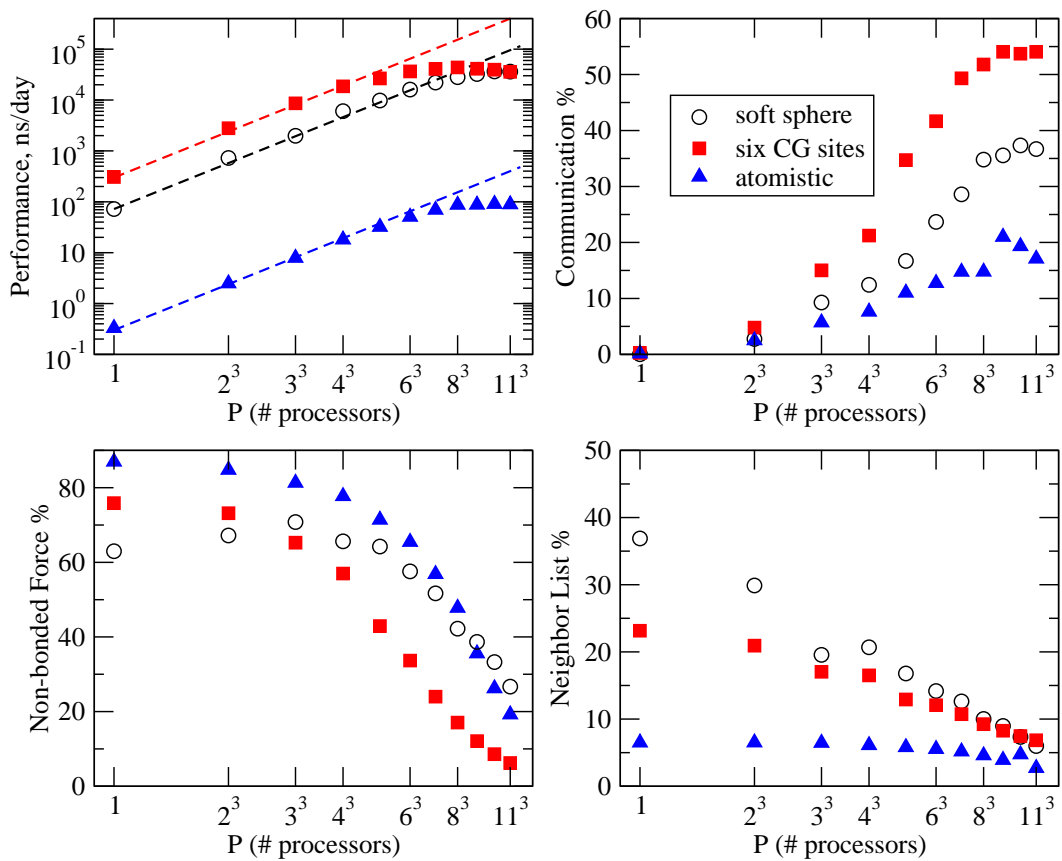}
\caption{Strong scaling performance and its detailed analysis consisting of communication, force calculations, and Verlet neighbor 
construction list for atomistic (blue triangles), soft sphere (black circles), 
and six CG site (red squares) models of polymers with degrees of polymerization of $300$:
the performance (nanosecond/day), percentage of communication, percentage of non-bonded pair forces computation, 
and percentage of Verlet neighbor list construction are reported for the atomistic and IECG simulations.
The dashed lines in the top left panel are the $100\%$ efficiency.
The simulation box lengths, $l_x$, are $345$, $194$, and $140$ \AA\ for the soft sphere, six CG site and atomistic models,
respectively. The statistical uncertainties are smaller than the symbol sizes for all molecular dynamics simulations.
}
\label{scaleN}
\end{figure}

In the case of multi-site models, particles can move to new processors and molecular connectivity information 
must be exchanged and updated between processors. The extra coding to manipulate the appropriate data structures 
and optimize the communication performance of the data exchange subtracts from the parallel efficiency of the
algorithm (see Figures \ref{scalability} and \ref{scaleN} and the discussion below).

To use the computational resources effectively and reduce the time to solve a problem, efficient strong scaling is required. 
Figure \ref{scalability} compares the strong scaling performance for the IECG simulations of polymer melts
with $N=300$ represented by soft sphere and six CG site models with the atomistic simulation performance,
which does not have considerable finite size effects for the structural and thermodynamical properties of interest.
All simulations were performed on the compute nodes of the Comet supercomputer, a high-performance computing 
cluster provided by Extreme Science and Engineering Discovery Environment (XSEDE) resources,\cite{xsede}
where the compute nodes consisted of Intel Xeon E5-2680v3 processors.

To minimize the communication in the spatial decomposition algorithm, the number of 
processors in each dimension was chosen to make each processor's box cubic.
The Verlet neighbor list was updated for each MD step for these sets of IECG simulations, 
which allowed us to use the spatial decomposition algorithm by adopting the appropriate CG timestep. 
The top panel presents the CPU strong scaling factor, when the performance is only reported in terms of MD step per second.   
Note that the dashed lines show $100\%$ parallel scalability (ideal scaling); 
therefore, excellent strong scaling is observed for the soft sphere 
and the atomistic models, while good strong scalability is shown for the six CG site model. 

Due to the significant increase in the range of the effective CG potential, 
the soft sphere model in comparison with the six CG site and the atomistic models requires more computation time to 
complete one MD step.
However, considering that a significantly larger CG timestep can be used for the soft sphere model
as compared with the six site and atomistic models,
the performance of soft sphere models considerably increases when the performance is 
reported in terms of ns per day. Note that due to the long-range effective CG potential 
for the soft sphere model, the simulation boxlength is the highest. 
In addition, larger CG timesteps can be used for serial runs as compared with the parallel runs due to the 
spatial decomposition algorithm and the enhanced dynamics of the CG system. 
For instance, it is possible that a bond between the CG sites becomes large: 
when a bond that straddles two subdomains becomes too large,
one of the two CG sites forming the bond is no longer within the surrounding ghost CG units
and the spatial decomposition algorithm practically fails.
Indeed, it was found that as the size of subdomain is increased,
the spatial decomposition algorithm can handle larger CG timesteps.

\clearpage

Figure \ref{scaleN} provides a detailed analysis of the parallel scalability for
the aforementioned IECG and atomistic simulations. 
In summary, it shows that the overall time consumption for a relatively short number of 
processors is due to the nonbonded force evaluations, but it switches to the communication 
for a relatively large number of processors. 
Considering the Message Passing Interface timing breakdown\cite{Plimpton1995}, 
the right top panel shows the
percentage of communication time as the number of processors increases. 
As the number of processors increase, the percentage of communication 
increases with the highest slope for the six site model most likely due to
the effective CG potential that is more long-ranged than the atomistic one
and the absence of molecular connectivity information in the soft sphere model. 
The range of potential in the atomistic simulations is much less than 
the IECG simulations, which results in lower percentage of communication for 
exchanging the particles from nearby boxes in the atomistic simulations.

The left bottom panel of Figure \ref{scaleN} shows that there is a crossover in the percentage of nonbonded force
calculations in MD simulations for the aforementioned models.
The percentage of nonbonded force calculations is almost a plateau for a relatively 
small number of processors, but the crossover occurs at a relatively large number of processors,
which is highest for the soft sphere model.
In addition, unlike the percentage of communication time, which almost
reaches a plateau for a large number of processors, the percentage of nonbonded force
computation time for all models does not show plateau behavior for a relatively large number of
processors.

The right bottom panel of Figure \ref{scaleN} shows the percentage of 
Verlet neighbor list calculation time, which is usually highest for the soft sphere model.
This is expected as the range of the effective CG potential and the number of pairwise interactions
are the highest for the soft sphere model. Similarly the six CG site model requires more percentage of 
consumption time for all processors as compared to the atomistic one, noting that the 
consumption time for the atomistic neighbor list calculations is almost 
constant for a relatively large number of processors.  

To end this section, it is worth mentioning that one can adjust the size and shape of processor 
sub-domains within the simulation box to balance the number of particles (load balance) 
more evenly across processors. The load balancing used in the above analysis was static
in the sense that one performs the balancing once, before MD simulations.
In addition, the analysis was done for relatively homogeneous systems, 
where uniform load balancing naturally emerges.
For the inhomogeneous systems, other approaches such as the Hilbert space-filling curve and 
cell task methods as discussed in the literature \cite{Grime2014,Brazdova2008,Meyer2014} may be used. 

\section{Conclusions}
\label{consec}

The IECG approach is a high level CG method, which allows modeling systems consisting 
of long chain polymers. It can be developed from a top-down or bottom-up approach. 
The bottom-up approach allows one to directly compare the computational 
efficiency of the IECG simulations with the atomistic simulations. 
Due to the nature of long-range effective IECG potential, 
relatively larger system sizes than the atomistic simulations are required
when computing relevant structural and thermodynamical properties.
On the basis of the data just discussed, it is clear that the 
selection of suitable parameters plays a pivotal role in controlling the computational 
efficiency of the IECG simulations. 

The recommendations from this work are to employ the
following parameters for the relevant IECG simulations at various state points:
(i) truncating the effective CG potential at distances, which
include the attractive part of the effective CG potential/force 
allows accurate and efficient IECG simulations; a cutoff distance of
$1.2 \: \sigma_{\rm{F1}}$ turns out to fulfill this requirement, where $\sigma_{\rm{F1}}$
is the first zero of the force over the range of distance greater than zero; 
(ii) considering the neighbor skin distance and the frequency of updating the Verlet neighbor list,
a timestep of $0.05$ of the shortest characteristic time, $\tau$, may be used for the IECG
simulations, where $\tau$ can be estimated from eq \ref{tausoft} and eq \ref{tausite}; 
(iii) the optimum neighbor skin distance depends on curvature of the effective CG potential, the CG timestep, 
and the frequency of updating the Verlet neighbor list;
as the level of CG changes for a given polymer melt, 
the curvature of the effective CG potential changes, which can affect the suitable neighbor skin distance; 
for a high level of CG the neighbor skin distances of $0.1 \: \sigma_{\rm{F1}}$ work reasonably well with the
standard frequency of the Verlet neighbor list update, but as the effective CG potential decays steeper for 
high resolution multi-site models relatively larger neighbor skin distances such as $0.4 \: \sigma_{\rm{F1}}$ are recommended;
note that the curvature of the effective CG potential, at relevant densities and temperatures and at the level of CG reported here, 
does not change considerably;
(iv) while the spatial decomposition algorithm can be used to speed up MD simulations using a large number of 
processors, the sub-domain size decreases, which can influence the selection of the suitable parameters; for instance,  
depending on the number of processors, smaller CG timesteps may be used for parallel MD simulations; 
(v) when the physical problem involves large correlation lengths associated 
with the properties of interest for long chain polymer melts, atomistic simulations,
which use relatively small system sizes, are more likely to be unsuccessful and it is 
inevitable to use relatively large system sizes to avoid finite size effects;
in these circumstances, a soft sphere model is perhaps the most efficient model to study
the global properties, considering that the free energy surface is smoothest.

The soft sphere and multi-site models in the IECG simulations allow one to treat the same molecular system at
different resolutions while offering millions of magnitude computational efficiency 
(as determined by eq \ref{Eqeff} and presented in Table \ref{tabeff} for various polymer melts). 
Considering the suitable MD parameters in the IECG and atomistic MD simulations,
the soft sphere models turn out to be computationally 
most efficient while the high-resolution multi-site models give more detailed descriptions of the structure on
the local scale.
The conclusions reached with regard to the optimal choice of the CG-MD parameters and the computational efficiency 
should also be applicable to other high level CG approaches.

\begin{acknowledgement}
This work was supported by the National Science Foundation (NSF) Grant No. CHE-13665466.
This work used the Extreme Science and Engineering Discovery Environment (XSEDE),\cite{xsede}
which is supported by the National Science Foundation Grant No. ACI-1548562.
This work used the XSEDE COMET at the San Diego Supercomputer Center through allocation
TG-CHE100082.
We thank Dr. David Ozog, and Prof. Allen D. Malony for useful discussions in the initial stage of this project.
We are grateful to Paula J. Seeger for reading/editing the manuscript.
\end{acknowledgement}

\setcitestyle{square}


\begin{thebibliography}{43}%

\makeatletter
\providecommand{\latin}[1]{#1}
\makeatletter
\providecommand{\doi}
  {\begingroup\let\do\@makeother\dospecials
  \catcode`\{=1 \catcode`\}=2 \doi@aux}
\providecommand{\doi@aux}[1]{\endgroup\texttt{#1}}
\makeatother
\providecommand*\natexlab[1]{#1}
\providecommand*\mciteSetBstSublistMode[1]{}
\providecommand*\mciteSetBstMaxWidthForm[2]{}
\providecommand*\mciteBstWouldAddEndPuncttrue
  {\def\EndOfBibitem{\unskip.}}
\providecommand*\mciteBstWouldAddEndPunctfalse
  {\let\EndOfBibitem\relax}
\providecommand*\mciteSetBstMidEndSepPunct[3]{}
\providecommand*\mciteSetBstSublistLabelBeginEnd[3]{}
\providecommand*\EndOfBibitem{}
\mciteSetBstSublistMode{f}
\mciteSetBstMaxWidthForm{subitem}{(\alph{mcitesubitemcount})}
\mciteSetBstSublistLabelBeginEnd
  {\mcitemaxwidthsubitemform\space}
  {\relax}
  {\relax}

\bibitem[Rahman and Stillinger(1971)Rahman, and Stillinger]{Rahman1971}
Rahman,~A.; Stillinger,~F.~H. {Molecular Dynamics Study of Liquid Water}.
  \emph{J. Chem. Phys.} \textbf{1971}, \emph{55}, 3336--3359\relax
\mciteBstWouldAddEndPuncttrue
\mciteSetBstMidEndSepPunct{\mcitedefaultmidpunct}
{\mcitedefaultendpunct}{\mcitedefaultseppunct}\relax
\EndOfBibitem
\bibitem[Allen and Tildesley(1987)Allen, and Tildesley]{Allen1987}
Allen,~M.~P.; Tildesley,~D.~J. \emph{{Computer Simulations of Liquids}}; Oxford
  University Press: Oxford, 1987\relax
\mciteBstWouldAddEndPuncttrue
\mciteSetBstMidEndSepPunct{\mcitedefaultmidpunct}
{\mcitedefaultendpunct}{\mcitedefaultseppunct}\relax
\EndOfBibitem
\bibitem[Frenkel and Smit(2002)Frenkel, and Smit]{Frenkel2002}
Frenkel,~D.; Smit,~B. \emph{{Understanding Molecular Simulations}}, 2nd ed.;
  Academic Press: San Diego, 2002\relax
\mciteBstWouldAddEndPuncttrue
\mciteSetBstMidEndSepPunct{\mcitedefaultmidpunct}
{\mcitedefaultendpunct}{\mcitedefaultseppunct}\relax
\EndOfBibitem
\bibitem[Salerno \latin{et~al.}(2016)Salerno, Agrawal, Perahia, and
  Grest]{Salerno2016}
Salerno,~K.~M.; Agrawal,~A.; Perahia,~D.; Grest,~G.~S. {Resolving Dynamic
  Properties of Polymers through Coarse-Grained Computational Studies}.
  \emph{Phys. Rev. Lett.} \textbf{2016}, \emph{116}, 3--7\relax
\mciteBstWouldAddEndPuncttrue
\mciteSetBstMidEndSepPunct{\mcitedefaultmidpunct}
{\mcitedefaultendpunct}{\mcitedefaultseppunct}\relax
\EndOfBibitem
\bibitem[Reith \latin{et~al.}(2003)Reith, P{\"{u}}tz, and
  M{\"{u}}ller-Plathe]{Reith2003}
Reith,~D.; P{\"{u}}tz,~M.; M{\"{u}}ller-Plathe,~F. {Deriving Effective
  Mesoscale Potentials from Atomistic Simulations}. \emph{J. Comp. Chem.}
  \textbf{2003}, \emph{24}, 1624--1636\relax
\mciteBstWouldAddEndPuncttrue
\mciteSetBstMidEndSepPunct{\mcitedefaultmidpunct}
{\mcitedefaultendpunct}{\mcitedefaultseppunct}\relax
\EndOfBibitem
\bibitem[Harmandaris \latin{et~al.}(2006)Harmandaris, Adhikari, {Van Der Vegt},
  and Kremer]{Harmandaris2006}
Harmandaris,~V.~A.; Adhikari,~N.~P.; {Van Der Vegt},~N. F.~A.; Kremer,~K.
  {Hierarchical Modeling of Polystyrene: From Atomistic to Coarse-Grained
  Simulations}. \emph{Macromolecules} \textbf{2006}, \emph{39},
  6708--6719\relax
\mciteBstWouldAddEndPuncttrue
\mciteSetBstMidEndSepPunct{\mcitedefaultmidpunct}
{\mcitedefaultendpunct}{\mcitedefaultseppunct}\relax
\EndOfBibitem
\bibitem[Guenza(2018)]{Guenza2018}
Guenza,~M.~G. \emph{Coarse-Grained Modeling of Biomolecules}; Taylor {\&}
  Francis Group, LLC; CRC Press: Boca Raton, 2018; Chapter 2, p~27\relax
\mciteBstWouldAddEndPuncttrue
\mciteSetBstMidEndSepPunct{\mcitedefaultmidpunct}
{\mcitedefaultendpunct}{\mcitedefaultseppunct}\relax
\EndOfBibitem
\bibitem[Clark \latin{et~al.}(2013)Clark, McCarty, and Guenza]{Clark2013}
Clark,~A.~J.; McCarty,~J.; Guenza,~M.~G. {Effective Potentials for Representing
  Polymers in Melts as Chains of Interacting Soft Particles}. \emph{J. Chem.
  Phys.} \textbf{2013}, \emph{139}, 124906\relax
\mciteBstWouldAddEndPuncttrue
\mciteSetBstMidEndSepPunct{\mcitedefaultmidpunct}
{\mcitedefaultendpunct}{\mcitedefaultseppunct}\relax
\EndOfBibitem
\bibitem[Dinpajooh and Guenza(2018)Dinpajooh, and Guenza]{Dinpajooh2017}
Dinpajooh,~M.; Guenza,~M.~G. {On the Density Dependence of the Integral
  Equation Coarse-Graining Effective Potential}. \emph{J. Phys. Chem. B}
  \textbf{2018}, \emph{122}, 3426--3440\relax
\mciteBstWouldAddEndPuncttrue
\mciteSetBstMidEndSepPunct{\mcitedefaultmidpunct}
{\mcitedefaultendpunct}{\mcitedefaultseppunct}\relax
\EndOfBibitem
\bibitem[Clark \latin{et~al.}(2012)Clark, Mccarty, Lyubimov, and
  Guenza]{Clark2012}
Clark,~A.~J.; Mccarty,~J.; Lyubimov,~I.~Y.; Guenza,~M.~G. {Thermodynamic
  Consistency in Variable-Level Coarse Graining of Polymeric Liquids}.
  \emph{Phys. Rev. Lett.} \textbf{2012}, \emph{109}, 168301\relax
\mciteBstWouldAddEndPuncttrue
\mciteSetBstMidEndSepPunct{\mcitedefaultmidpunct}
{\mcitedefaultendpunct}{\mcitedefaultseppunct}\relax
\EndOfBibitem
\bibitem[McCarty \latin{et~al.}(2014)McCarty, Clark, Copperman, and
  Guenza]{McCarty2014}
McCarty,~J.; Clark,~A.~J.; Copperman,~J.; Guenza,~M.~G. {An Analytical
  Coarse-Graining Method which Preserves the Free Energy, Structural
  Correlations, and Thermodynamic State of Polymer Melts from the Atomistic to
  the Mesoscale}. \emph{J. Chem. Phys.} \textbf{2014}, \emph{140}, 204913\relax
\mciteBstWouldAddEndPuncttrue
\mciteSetBstMidEndSepPunct{\mcitedefaultmidpunct}
{\mcitedefaultendpunct}{\mcitedefaultseppunct}\relax
\EndOfBibitem
\bibitem[Dinpajooh and Guenza(2017)Dinpajooh, and Guenza]{Dinpajooh2017a}
Dinpajooh,~M.; Guenza,~M.~G. {Thermodynamic Consistency in the Structure-based
  Integral Equation Coarse-Grained Method}. \emph{Polymer} \textbf{2017},
  \emph{117}, 282--286\relax
\mciteBstWouldAddEndPuncttrue
\mciteSetBstMidEndSepPunct{\mcitedefaultmidpunct}
{\mcitedefaultendpunct}{\mcitedefaultseppunct}\relax
\EndOfBibitem
\bibitem[Yatsenko \latin{et~al.}(2004)Yatsenko, Sambriski, Nemirovskaya, and
  Guenza]{Yatsenko2004}
Yatsenko,~G.; Sambriski,~E.~J.; Nemirovskaya,~M.~A.; Guenza,~M. {Analytical
  Soft-Core Potentials for Macromolecular Fluids and Mixtures}. \emph{Phys.
  Rev. Lett.} \textbf{2004}, \emph{93}, 257803\relax
\mciteBstWouldAddEndPuncttrue
\mciteSetBstMidEndSepPunct{\mcitedefaultmidpunct}
{\mcitedefaultendpunct}{\mcitedefaultseppunct}\relax
\EndOfBibitem
\bibitem[Clark and Guenza(2010)Clark, and Guenza]{Clark2010}
Clark,~A.~J.; Guenza,~M.~G. {Mapping of Polymer Melts onto Liquids of
  Soft-Colloidal Chains}. \emph{J. Chem. Phys.} \textbf{2010}, \emph{132},
  044902\relax
\mciteBstWouldAddEndPuncttrue
\mciteSetBstMidEndSepPunct{\mcitedefaultmidpunct}
{\mcitedefaultendpunct}{\mcitedefaultseppunct}\relax
\EndOfBibitem
\bibitem[McCarty \latin{et~al.}(2012)McCarty, Clark, Lyubimov, and
  Guenza]{McCarty2012}
McCarty,~J.; Clark,~A.~J.; Lyubimov,~I.~Y.; Guenza,~M.~G. {Thermodynamic
  Consistency between Analytic Integral Equation Theory and Coarse-Grained
  Molecular Dynamics Simulations of Homopolymer Melts}. \emph{Macromolecules}
  \textbf{2012}, \emph{45}, 8482--8493\relax
\mciteBstWouldAddEndPuncttrue
\mciteSetBstMidEndSepPunct{\mcitedefaultmidpunct}
{\mcitedefaultendpunct}{\mcitedefaultseppunct}\relax
\EndOfBibitem
\bibitem[Carnahan and Starling(1970)Carnahan, and Starling]{Carnahan1970}
Carnahan,~N.~F.; Starling,~K.~E. {Thermodynamic Properties of a Rigid-Sphere
  Fluid}. \emph{J. Chem. Phys.} \textbf{1970}, \emph{53}, 600--603\relax
\mciteBstWouldAddEndPuncttrue
\mciteSetBstMidEndSepPunct{\mcitedefaultmidpunct}
{\mcitedefaultendpunct}{\mcitedefaultseppunct}\relax
\EndOfBibitem
\bibitem[Carbone \latin{et~al.}(2008)Carbone, Varzaneh, Chen, and
  M{\"{u}}ller-Plathe]{Carbone2008}
Carbone,~P.; Varzaneh,~H. A.~K.; Chen,~X.; M{\"{u}}ller-Plathe,~F.
  {Transferability of Coarse-Grained Force Fields: The Polymer Case}. \emph{J.
  Chem. Phys.} \textbf{2008}, \emph{128}, 064904\relax
\mciteBstWouldAddEndPuncttrue
\mciteSetBstMidEndSepPunct{\mcitedefaultmidpunct}
{\mcitedefaultendpunct}{\mcitedefaultseppunct}\relax
\EndOfBibitem
\bibitem[Harmandaris \latin{et~al.}(2003)Harmandaris, Mavrantzas, Theodorou,
  Kr{\"{o}}ger, Ram{\'{i}}rez, Ottinger, and Vlassopoulos]{Harmandaris2003}
Harmandaris,~V.~A.; Mavrantzas,~V.~G.; Theodorou,~D.~N.; Kr{\"{o}}ger,~M.;
  Ram{\'{i}}rez,~J.; Ottinger,~H.~C.; Vlassopoulos,~D. {Crossover from the
  Rouse to the Entangled Polymer Melt Regime: Signals from Long, Detailed
  Atomistic Molecular Dynamics Simulations, Supported by Rheological
  Experiments}. \emph{Macromolecules} \textbf{2003}, \emph{36},
  1376--1387\relax
\mciteBstWouldAddEndPuncttrue
\mciteSetBstMidEndSepPunct{\mcitedefaultmidpunct}
{\mcitedefaultendpunct}{\mcitedefaultseppunct}\relax
\EndOfBibitem
\bibitem[Das \latin{et~al.}(2012)Das, Lu, Andersen, and Voth]{Das2012}
Das,~A.; Lu,~L.; Andersen,~H.~C.; Voth,~G.~A. {The Multiscale Coarse-Graining
  Method. X. Improved Algorithms for Constructing Coarse-grained Potentials for
  Molecular Systems}. \emph{J. Chem. Phys.} \textbf{2012}, \emph{136},
  194114\relax
\mciteBstWouldAddEndPuncttrue
\mciteSetBstMidEndSepPunct{\mcitedefaultmidpunct}
{\mcitedefaultendpunct}{\mcitedefaultseppunct}\relax
\EndOfBibitem
\bibitem[Ramos \latin{et~al.}(2015)Ramos, Vega, and
  Martinez-Salazar]{Ramos2015}
Ramos,~J.; Vega,~J.~F.; Martinez-Salazar,~J. {Molecular Dynamics Simulations
  for the Description of Experimental Molecular Conformation, Melt Dynamics,
  and Phase Transitions in Polyethylene}. \emph{Macromolecules} \textbf{2015},
  \emph{48}, 5016--5027\relax
\mciteBstWouldAddEndPuncttrue
\mciteSetBstMidEndSepPunct{\mcitedefaultmidpunct}
{\mcitedefaultendpunct}{\mcitedefaultseppunct}\relax
\EndOfBibitem
\bibitem[Cao and Voth(2015)Cao, and Voth]{Cao2015}
Cao,~Z.; Voth,~G.~A. {The Multiscale Coarse-Graining Method. XI. Accurate
  Interactions Based on the Centers of Charge of Coarse-grained Sites}.
  \emph{J. Chem. Phys.} \textbf{2015}, \emph{143}, 243116\relax
\mciteBstWouldAddEndPuncttrue
\mciteSetBstMidEndSepPunct{\mcitedefaultmidpunct}
{\mcitedefaultendpunct}{\mcitedefaultseppunct}\relax
\EndOfBibitem
\bibitem[Izvekov and Voth(2005)Izvekov, and Voth]{Izvekov2005}
Izvekov,~S.; Voth,~G.~A. {Multiscale Coarse Graining of Liquid-State Systems}.
  \emph{J. Chem. Phys.} \textbf{2005}, \emph{123}, 134105\relax
\mciteBstWouldAddEndPuncttrue
\mciteSetBstMidEndSepPunct{\mcitedefaultmidpunct}
{\mcitedefaultendpunct}{\mcitedefaultseppunct}\relax
\EndOfBibitem
\bibitem[Shell(2008)]{Shell2008}
Shell,~M.~S. {The Relative Entropy is Fundamental to Multiscale and Inverse
  Thermodynamic Problems}. \emph{J. Chem. Phys.} \textbf{2008}, \emph{129},
  144108\relax
\mciteBstWouldAddEndPuncttrue
\mciteSetBstMidEndSepPunct{\mcitedefaultmidpunct}
{\mcitedefaultendpunct}{\mcitedefaultseppunct}\relax
\EndOfBibitem
\bibitem[Brini \latin{et~al.}(2013)Brini, Algaer, Ganguly, Li,
  Rodr{\'{i}}guez-Ropero, and van~der Vegt]{Brini2013}
Brini,~E.; Algaer,~E.~A.; Ganguly,~P.; Li,~C.; Rodr{\'{i}}guez-Ropero,~F.;
  van~der Vegt,~N. F.~A. {Systematic Coarse-Graining Methods for Soft Matter
  Simulations - a Review}. \emph{Soft Matter} \textbf{2013}, \emph{9},
  2108--2119\relax
\mciteBstWouldAddEndPuncttrue
\mciteSetBstMidEndSepPunct{\mcitedefaultmidpunct}
{\mcitedefaultendpunct}{\mcitedefaultseppunct}\relax
\EndOfBibitem
\bibitem[Dunn and Noid(2015)Dunn, and Noid]{Dunn2015}
Dunn,~N.~J.; Noid,~W.~G. {Bottom-up Coarse-Grained Models that Accurately
  Describe the Structure, Pressure, and Compressibility of Molecular Liquids}.
  \emph{J. Chem. Phys.} \textbf{2015}, \emph{143}, 243148\relax
\mciteBstWouldAddEndPuncttrue
\mciteSetBstMidEndSepPunct{\mcitedefaultmidpunct}
{\mcitedefaultendpunct}{\mcitedefaultseppunct}\relax
\EndOfBibitem
\bibitem[Rudzinski and Noid(2015)Rudzinski, and Noid]{Rudzinski2015}
Rudzinski,~J.~F.; Noid,~W.~G. {Bottom-up Coarse-Graining of Peptide Ensembles
  and Helix-coil Transitions}. \emph{J. Chem. Theory Comput.} \textbf{2015},
  \emph{11}, 1278--1291\relax
\mciteBstWouldAddEndPuncttrue
\mciteSetBstMidEndSepPunct{\mcitedefaultmidpunct}
{\mcitedefaultendpunct}{\mcitedefaultseppunct}\relax
\EndOfBibitem
\bibitem[Avenda{\~{n}}o \latin{et~al.}(2011)Avenda{\~{n}}o, Lafitte, Galindo,
  Adjiman, Jackson, and M{\"{u}}ller]{Avendano2011}
Avenda{\~{n}}o,~C.; Lafitte,~T.; Galindo,~A.; Adjiman,~C.~S.; Jackson,~G.;
  M{\"{u}}ller,~E.~A. {SAFT-$\gamma$ Force Field for the Simulation of
  Molecular Fluids. 1. A Single-site Coarse grained Model of Carbon Dioxide}.
  \emph{J. Phys. Chem. B} \textbf{2011}, \emph{115}, 11154--11169\relax
\mciteBstWouldAddEndPuncttrue
\mciteSetBstMidEndSepPunct{\mcitedefaultmidpunct}
{\mcitedefaultendpunct}{\mcitedefaultseppunct}\relax
\EndOfBibitem
\bibitem[Gil-Villegas \latin{et~al.}(1997)Gil-Villegas, Galindo, Whitehead,
  Mills, Jackson, and Burgess]{Gil-Villegas1997}
Gil-Villegas,~A.; Galindo,~A.; Whitehead,~P.~J.; Mills,~S.~J.; Jackson,~G.;
  Burgess,~A.~N. {Statistical Associating Fluid Theory for Chain Molecules with
  Attractive Potentials of Variable Range}. \emph{J. Chem. Phys.}
  \textbf{1997}, \emph{106}, 4168--4186\relax
\mciteBstWouldAddEndPuncttrue
\mciteSetBstMidEndSepPunct{\mcitedefaultmidpunct}
{\mcitedefaultendpunct}{\mcitedefaultseppunct}\relax
\EndOfBibitem
\bibitem[Hsu \latin{et~al.}(2015)Hsu, Xia, Arturo, and Keten]{Hsu2015}
Hsu,~D.~D.; Xia,~W.; Arturo,~S.~G.; Keten,~S. {Thermomechanically Consistent
  and Temperature Transferable Coarse-Graining of Atactic Polystyrene}.
  \emph{Macromolecules} \textbf{2015}, \emph{48}, 3057--3068\relax
\mciteBstWouldAddEndPuncttrue
\mciteSetBstMidEndSepPunct{\mcitedefaultmidpunct}
{\mcitedefaultendpunct}{\mcitedefaultseppunct}\relax
\EndOfBibitem
\bibitem[Kremer and Grest(1990)Kremer, and Grest]{Kremer1990}
Kremer,~K.; Grest,~G.~S. {Dynamics of Entangled Linear Polymer Melts: A
  Molecular-Dynamics Simulation}. \emph{J. Chem. Phys.} \textbf{1990},
  \emph{92}, 5057--5086\relax
\mciteBstWouldAddEndPuncttrue
\mciteSetBstMidEndSepPunct{\mcitedefaultmidpunct}
{\mcitedefaultendpunct}{\mcitedefaultseppunct}\relax
\EndOfBibitem
\bibitem[Fritz \latin{et~al.}(2009)Fritz, Herbers, Kremer, and van~der
  Vegt]{Fritz2009}
Fritz,~D.; Herbers,~C.~R.; Kremer,~K.; van~der Vegt,~N. F.~A. {Hierarchical
  Modeling of Polymer Permeation}. \emph{Soft Matter} \textbf{2009}, \emph{5},
  4556\relax
\mciteBstWouldAddEndPuncttrue
\mciteSetBstMidEndSepPunct{\mcitedefaultmidpunct}
{\mcitedefaultendpunct}{\mcitedefaultseppunct}\relax
\EndOfBibitem
\bibitem[Lyubimov \latin{et~al.}(2010)Lyubimov, McCarty, Clark, and
  Guenza]{Lyubimov2010}
Lyubimov,~I.~Y.; McCarty,~J.; Clark,~A.; Guenza,~M.~G. {Analytical Rescaling of
  Polymer Dynamics from Mesoscale Simulations}. \emph{J. Chem. Phys.}
  \textbf{2010}, \emph{132}, 224903\relax
\mciteBstWouldAddEndPuncttrue
\mciteSetBstMidEndSepPunct{\mcitedefaultmidpunct}
{\mcitedefaultendpunct}{\mcitedefaultseppunct}\relax
\EndOfBibitem
\bibitem[Lyubimov and Guenza(2011)Lyubimov, and Guenza]{Lyubimov2011}
Lyubimov,~I.; Guenza,~M.~G. {First-principle Approach to Rescale the Dynamics
  of Simulated Coarse-Grained Macromolecular Liquids}. \emph{Phys. Rev. E}
  \textbf{2011}, \emph{84}, 16--18\relax
\mciteBstWouldAddEndPuncttrue
\mciteSetBstMidEndSepPunct{\mcitedefaultmidpunct}
{\mcitedefaultendpunct}{\mcitedefaultseppunct}\relax
\EndOfBibitem
\bibitem[Fritz \latin{et~al.}(2011)Fritz, Koschke, Harmandaris, van~der Vegt,
  and Kremer]{Fritz2011}
Fritz,~D.; Koschke,~K.; Harmandaris,~V.~A.; van~der Vegt,~N. F.~A.; Kremer,~K.
  {Multiscale Modeling of Soft Matter: Scaling of Dynamics}. \emph{Phys. Chem.
  Chem. Phys.} \textbf{2011}, \emph{13}, 10412\relax
\mciteBstWouldAddEndPuncttrue
\mciteSetBstMidEndSepPunct{\mcitedefaultmidpunct}
{\mcitedefaultendpunct}{\mcitedefaultseppunct}\relax
\EndOfBibitem
\bibitem[Lyubimov and Guenza(2013)Lyubimov, and Guenza]{Lyubimov2013}
Lyubimov,~I.~Y.; Guenza,~M.~G. {Theoretical Reconstruction of Realistic
  Dynamics of Highly Coarse-Grained cis-1,4-Polybutadiene Melts}. \emph{J.
  Chem. Phys.} \textbf{2013}, \emph{138}, 12A546\relax
\mciteBstWouldAddEndPuncttrue
\mciteSetBstMidEndSepPunct{\mcitedefaultmidpunct}
{\mcitedefaultendpunct}{\mcitedefaultseppunct}\relax
\EndOfBibitem
\bibitem[Davtyan \latin{et~al.}(2016)Davtyan, Voth, and Andersen]{Davtyan2016}
Davtyan,~A.; Voth,~G.~A.; Andersen,~H.~C. {Dynamic Force Matching: Construction
  of Dynamic Coarse-Grained Models with Realistic Short Time Dynamics and
  Accurate Long Time Dynamics}. \emph{J. Chem. Phys.} \textbf{2016},
  \emph{145}, 224107\relax
\mciteBstWouldAddEndPuncttrue
\mciteSetBstMidEndSepPunct{\mcitedefaultmidpunct}
{\mcitedefaultendpunct}{\mcitedefaultseppunct}\relax
\EndOfBibitem
\bibitem[Plimpton(1995)]{Plimpton1995}
Plimpton,~S. {Fast Parallel Algorithms for Short-Range Molecular Dynamics}.
  \emph{J. Comp. Phys.} \textbf{1995}, \emph{117}, 1--19\relax
\mciteBstWouldAddEndPuncttrue
\mciteSetBstMidEndSepPunct{\mcitedefaultmidpunct}
{\mcitedefaultendpunct}{\mcitedefaultseppunct}\relax
\EndOfBibitem
\bibitem[Bussi \latin{et~al.}(2007)Bussi, Donadio, and Parrinello]{Bussi2007}
Bussi,~G.; Donadio,~D.; Parrinello,~M. {Canonical Sampling Through Velocity
  Rescaling}. \emph{J. Chem. Phys.} \textbf{2007}, \emph{126}, 014101\relax
\mciteBstWouldAddEndPuncttrue
\mciteSetBstMidEndSepPunct{\mcitedefaultmidpunct}
{\mcitedefaultendpunct}{\mcitedefaultseppunct}\relax
\EndOfBibitem
\bibitem[Halverson \latin{et~al.}(2013)Halverson, Brandes, Lenz, Arnold, Bevc,
  Starchenko, Kremer, Stuehn, and Reith]{Halverson2013}
Halverson,~J.~D.; Brandes,~T.; Lenz,~O.; Arnold,~A.; Bevc,~S.; Starchenko,~V.;
  Kremer,~K.; Stuehn,~T.; Reith,~D. {ESPResSo++: A Modern Multiscale Simulation
  Package for Soft Matter Systems}. \emph{Comput. Phys. Commun.} \textbf{2013},
  \emph{184}, 1129--1149\relax
\mciteBstWouldAddEndPuncttrue
\mciteSetBstMidEndSepPunct{\mcitedefaultmidpunct}
{\mcitedefaultendpunct}{\mcitedefaultseppunct}\relax
\EndOfBibitem
\bibitem[Verlet(1967)]{Verlet1967}
Verlet,~L. {Computer "Experiment" on Classical Fluids. I. Thermodynamical
  Properties of Lennard-Jones Molecules}. \emph{Phys. Rev.} \textbf{1967},
  \emph{159}, 98\relax
\mciteBstWouldAddEndPuncttrue
\mciteSetBstMidEndSepPunct{\mcitedefaultmidpunct}
{\mcitedefaultendpunct}{\mcitedefaultseppunct}\relax
\EndOfBibitem
\bibitem[Chialvo and Debenedetti(1990)Chialvo, and Debenedetti]{Chialvo1990}
Chialvo,~A.~A.; Debenedetti,~P.~G. {On the Use of the Verlet Neighbor List in
  Molecular Dynamics}. \emph{Comput. Phys. Commun.} \textbf{1990}, \emph{60},
  215--224\relax
\mciteBstWouldAddEndPuncttrue
\mciteSetBstMidEndSepPunct{\mcitedefaultmidpunct}
{\mcitedefaultendpunct}{\mcitedefaultseppunct}\relax
\EndOfBibitem
\bibitem[Schweizer and Gurro(1997)Schweizer, and Gurro]{Schweizer1997}
Schweizer,~K.~S.; Gurro,~J.~G. {Integral Equation Theories of the Structure,
  Thermodynamics, and Phase Transitions of Polymer Fluids}. \emph{Adv. Chem.
  Phys.} \textbf{1997}, \emph{98}, 1--142\relax
\mciteBstWouldAddEndPuncttrue
\mciteSetBstMidEndSepPunct{\mcitedefaultmidpunct}
{\mcitedefaultendpunct}{\mcitedefaultseppunct}\relax
\EndOfBibitem
\bibitem[Stanley(1987)]{Stanley1987}
Stanley,~H.~E. \emph{{Introduction to Phase Transitions and Critical
  Phenomena}}; Oxford University Press: New York, 1987\relax
\mciteBstWouldAddEndPuncttrue
\mciteSetBstMidEndSepPunct{\mcitedefaultmidpunct}
{\mcitedefaultendpunct}{\mcitedefaultseppunct}\relax
\EndOfBibitem
\bibitem[Stukan \latin{et~al.}(2002)Stukan, Ivanov, Muller, Paul, and
  Binder]{Stukan2002}
Stukan,~M.~R.; Ivanov,~V.~A.; Muller,~M.; Paul,~W.; Binder,~K. {Finite Size
  Effects in Pressure Measurements for Monte Carlo Simulations of Lattice
  Polymer Models}. \emph{J. Chem. Phys.} \textbf{2002}, \emph{117},
  9934--9941\relax
\mciteBstWouldAddEndPuncttrue
\mciteSetBstMidEndSepPunct{\mcitedefaultmidpunct}
{\mcitedefaultendpunct}{\mcitedefaultseppunct}\relax
\EndOfBibitem
\bibitem[McCarty \latin{et~al.}(2010)McCarty, Lyubimov, and
  Guenza]{McCarty2010}
McCarty,~J.; Lyubimov,~I.~Y.; Guenza,~M.~G. {Effective Soft-Core Potentials and
  Mesoscopic Simulations of Binary Polymer Mixtures}. \emph{Macromolecules}
  \textbf{2010}, \emph{43}, 3964--3979\relax
\mciteBstWouldAddEndPuncttrue
\mciteSetBstMidEndSepPunct{\mcitedefaultmidpunct}
{\mcitedefaultendpunct}{\mcitedefaultseppunct}\relax
\EndOfBibitem
\bibitem[Hansen and McDonald(2003)Hansen, and McDonald]{Hansen2003}
Hansen,~J.~P.; McDonald,~I.~R. \emph{{Theory of Simple Liquids}}; Academic
  Press: Amsterdam, 2003\relax
\mciteBstWouldAddEndPuncttrue
\mciteSetBstMidEndSepPunct{\mcitedefaultmidpunct}
{\mcitedefaultendpunct}{\mcitedefaultseppunct}\relax
\EndOfBibitem
\bibitem[Schweizer and Curro(1987)Schweizer, and Curro]{Schweizer1987}
Schweizer,~K.~S.; Curro,~J.~G. {Integral-Equation Theory of the Structure of
  Polymer Melts}. \emph{Phys. Rev. Lett.} \textbf{1987}, \emph{58},
  246--249\relax
\mciteBstWouldAddEndPuncttrue
\mciteSetBstMidEndSepPunct{\mcitedefaultmidpunct}
{\mcitedefaultendpunct}{\mcitedefaultseppunct}\relax
\EndOfBibitem
\bibitem[Schweizer and Curro(1990)Schweizer, and Curro]{Schweizer1990}
Schweizer,~K.~S.; Curro,~J.~G. {RISM Theory of Polymer Liquids: Analytical
  Results for Continuum Models of Melts and Alloys}. \emph{Chem. Phys.}
  \textbf{1990}, \emph{149}, 105--127\relax
\mciteBstWouldAddEndPuncttrue
\mciteSetBstMidEndSepPunct{\mcitedefaultmidpunct}
{\mcitedefaultendpunct}{\mcitedefaultseppunct}\relax
\EndOfBibitem
\bibitem[Chandler and Andersen(1972)Chandler, and Andersen]{Chandler1972}
Chandler,~D.; Andersen,~H.~C. {Optimized Cluster Expansions for Classical
  Fluids. II. Theory of Molecular Liquids}. \emph{J. Chem. Phys.}
  \textbf{1972}, \emph{57}, 1930--1937\relax
\mciteBstWouldAddEndPuncttrue
\mciteSetBstMidEndSepPunct{\mcitedefaultmidpunct}
{\mcitedefaultendpunct}{\mcitedefaultseppunct}\relax
\EndOfBibitem
\bibitem[Guenza and Schweizer(1997)Guenza, and Schweizer]{Guenza1997}
Guenza,~M.; Schweizer,~K.~S. {Fluctuations Effects in Diblock Copolymer Fluids:
  Comparison of Theories and Experiment}. \emph{J. Chem. Phys.} \textbf{1997},
  \emph{106}, 7391--7410\relax
\mciteBstWouldAddEndPuncttrue
\mciteSetBstMidEndSepPunct{\mcitedefaultmidpunct}
{\mcitedefaultendpunct}{\mcitedefaultseppunct}\relax
\EndOfBibitem
\bibitem[Yatsenko \latin{et~al.}(2005)Yatsenko, Sambriski, and
  Guenza]{Yatsenko2005}
Yatsenko,~G.; Sambriski,~E.~J.; Guenza,~M.~G. {Coarse-grained Description of
  Polymer Blends as Interacting Soft-Colloidal Particles}. \emph{J. Chem.
  Phys.} \textbf{2005}, \emph{122}, 054907\relax
\mciteBstWouldAddEndPuncttrue
\mciteSetBstMidEndSepPunct{\mcitedefaultmidpunct}
{\mcitedefaultendpunct}{\mcitedefaultseppunct}\relax
\EndOfBibitem
\bibitem[Sankar and Tripathy(2015)Sankar, and Tripathy]{Sankar2015}
Sankar,~U.~K.; Tripathy,~M. {Dispersion, Depletion, and Bridging of Athermal
  and Attractive Nanorods in Polymer Melt}. \emph{Macromolecules}
  \textbf{2015}, \emph{48}, 432--442\relax
\mciteBstWouldAddEndPuncttrue
\mciteSetBstMidEndSepPunct{\mcitedefaultmidpunct}
{\mcitedefaultendpunct}{\mcitedefaultseppunct}\relax
\EndOfBibitem
\bibitem[Louis \latin{et~al.}(2000)Louis, Bolhuis, Hansen, and
  Meijer]{Louis2000}
Louis,~A.~A.; Bolhuis,~P.~G.; Hansen,~J.~P.; Meijer,~E.~J. {Can Polymer Coils
  Be Modeled as "Soft Colloids"?} \emph{Phys. Rev. Lett.} \textbf{2000},
  \emph{85}, 2522--2525\relax
\mciteBstWouldAddEndPuncttrue
\mciteSetBstMidEndSepPunct{\mcitedefaultmidpunct}
{\mcitedefaultendpunct}{\mcitedefaultseppunct}\relax
\EndOfBibitem
\bibitem[Dinpajooh and Guenza()Dinpajooh, and Guenza]{IECGweb}
Dinpajooh,~M.; Guenza,~M.~G. {The Integral Equation Coarse-Graining Method}.
  \url{https://iecgsim.uoregon.edu}\relax
\mciteBstWouldAddEndPuncttrue
\mciteSetBstMidEndSepPunct{\mcitedefaultmidpunct}
{\mcitedefaultendpunct}{\mcitedefaultseppunct}\relax
\EndOfBibitem
\bibitem[McCarty \latin{et~al.}(2009)McCarty, Lyubimov, and
  Guenza]{McCarty2009}
McCarty,~J.; Lyubimov,~I.~Y.; Guenza,~M.~G. {Multiscale Modeling of
  Coarse-Grained Macromolecular Liquids}. \emph{J. Phys. Chem. B}
  \textbf{2009}, \emph{113}, 11876--11886\relax
\mciteBstWouldAddEndPuncttrue
\mciteSetBstMidEndSepPunct{\mcitedefaultmidpunct}
{\mcitedefaultendpunct}{\mcitedefaultseppunct}\relax
\EndOfBibitem
\bibitem[Martin and Siepmann(1998)Martin, and Siepmann]{Martin1998}
Martin,~M.~G.; Siepmann,~J.~I. {Transferable Potentials for Phase Equilibria.
  1. United-Atom Description of n -Alkanes}. \emph{J. Phys. Chem. B}
  \textbf{1998}, \emph{102}, 2569--2577\relax
\mciteBstWouldAddEndPuncttrue
\mciteSetBstMidEndSepPunct{\mcitedefaultmidpunct}
{\mcitedefaultendpunct}{\mcitedefaultseppunct}\relax
\EndOfBibitem
\bibitem[Mei \latin{et~al.}(1991)Mei, Davenport, and Fernando]{Mei1991}
Mei,~J.; Davenport,~J.~W.; Fernando,~G.~W. {Analytic Embedded-atom Potentials
  for FCC Metals: Application to Liquid and Solid Copper}. \emph{Phys. Rev. B.}
  \textbf{1991}, \emph{43}, 4653--4658\relax
\mciteBstWouldAddEndPuncttrue
\mciteSetBstMidEndSepPunct{\mcitedefaultmidpunct}
{\mcitedefaultendpunct}{\mcitedefaultseppunct}\relax
\EndOfBibitem
\bibitem[Meyer(2014)]{Meyer2014}
Meyer,~R. {Efficient Parallelization of Molecular Dynamics Simulations with
  Short-Ranged Forces}. \emph{J. Phys. Conf. Ser.} \textbf{2014}, \emph{540},
  012006\relax
\mciteBstWouldAddEndPuncttrue
\mciteSetBstMidEndSepPunct{\mcitedefaultmidpunct}
{\mcitedefaultendpunct}{\mcitedefaultseppunct}\relax
\EndOfBibitem
\bibitem[Paul \latin{et~al.}(1995)Paul, Yoon, and Smith]{Paul1995}
Paul,~W.; Yoon,~D.~Y.; Smith,~G.~D. {An Optimized United Atom Model for
  Simulations of Polymethylene Melts}. \emph{J. Chem. Phys.} \textbf{1995},
  \emph{103}, 1702--1709\relax
\mciteBstWouldAddEndPuncttrue
\mciteSetBstMidEndSepPunct{\mcitedefaultmidpunct}
{\mcitedefaultendpunct}{\mcitedefaultseppunct}\relax
\EndOfBibitem
\bibitem[Marrink \latin{et~al.}(2007)Marrink, Risselada, Yefimov, Tieleman, and
  {De Vries}]{Marrink2007}
Marrink,~S.~J.; Risselada,~H.~J.; Yefimov,~S.; Tieleman,~D.~P.; {De
  Vries},~A.~H. {The MARTINI Force Field: Coarse-Grained Model for Biomolecular
  Simulations}. \emph{J. Phys. Chem. B} \textbf{2007}, \emph{111},
  7812--7824\relax
\mciteBstWouldAddEndPuncttrue
\mciteSetBstMidEndSepPunct{\mcitedefaultmidpunct}
{\mcitedefaultendpunct}{\mcitedefaultseppunct}\relax
\EndOfBibitem
\bibitem[M{\"{u}}ller-Plathe(2002)]{Muller-Plathe2002}
M{\"{u}}ller-Plathe,~F. {Coarse-graining in Polymer Simulation: From the
  Atomistic to the Mesoscopic Scale and Back}. \emph{ChemPhysChem}
  \textbf{2002}, \emph{3}, 754--769\relax
\mciteBstWouldAddEndPuncttrue
\mciteSetBstMidEndSepPunct{\mcitedefaultmidpunct}
{\mcitedefaultendpunct}{\mcitedefaultseppunct}\relax
\EndOfBibitem
\bibitem[Siu \latin{et~al.}(2012)Siu, Pluhackova, and B{\"{o}}ckmann]{Siu2012}
Siu,~S. W.~I.; Pluhackova,~K.; B{\"{o}}ckmann,~R.~A. {Optimization of the
  OPLS-AA Force Field for Long Hydrocarbons}. \emph{J. Chem. Theory Comput.}
  \textbf{2012}, \emph{8}, 1459--1470\relax
\mciteBstWouldAddEndPuncttrue
\mciteSetBstMidEndSepPunct{\mcitedefaultmidpunct}
{\mcitedefaultendpunct}{\mcitedefaultseppunct}\relax
\EndOfBibitem
\bibitem[Tuckerman \latin{et~al.}(1992)Tuckerman, Berne, and
  Martyna]{Tuckerman1992}
Tuckerman,~M.; Berne,~B.~J.; Martyna,~G.~J. {Reversible Multiple Time Scale
  Molecular Dynamics}. \emph{J. Chem. Phys.} \textbf{1992}, \emph{97},
  1990--2001\relax
\mciteBstWouldAddEndPuncttrue
\mciteSetBstMidEndSepPunct{\mcitedefaultmidpunct}
{\mcitedefaultendpunct}{\mcitedefaultseppunct}\relax
\EndOfBibitem
\bibitem[Marrink \latin{et~al.}(2004)Marrink, de~Vries, and Mark]{Marrink2004}
Marrink,~S.~J.; de~Vries,~A.~H.; Mark,~A.~E. {Coarse Grained Model for
  Semiquantitative Lipid Simulations}. \emph{J. Phys. Chem. B} \textbf{2004},
  \emph{108}, 750--760\relax
\mciteBstWouldAddEndPuncttrue
\mciteSetBstMidEndSepPunct{\mcitedefaultmidpunct}
{\mcitedefaultendpunct}{\mcitedefaultseppunct}\relax
\EndOfBibitem
\bibitem[Murtola \latin{et~al.}(2009)Murtola, Bunker, Vattulainen, Deserno, and
  Karttunen]{Murtola2009}
Murtola,~T.; Bunker,~A.; Vattulainen,~I.; Deserno,~M.; Karttunen,~M. {On Using
  a Too Large Integration Time Step in Molecular Dynamics Simulations of
  Coarse-Grained Molecular Models}. \emph{Phys. Chem. Chem. Phys.}
  \textbf{2009}, \emph{11}, 1869\relax
\mciteBstWouldAddEndPuncttrue
\mciteSetBstMidEndSepPunct{\mcitedefaultmidpunct}
{\mcitedefaultendpunct}{\mcitedefaultseppunct}\relax
\EndOfBibitem
\bibitem[Marrink \latin{et~al.}(2010)Marrink, Periole, Tieleman, and
  de~Vries]{Marrink2010a}
Marrink,~S.~J.; Periole,~X.; Tieleman,~D.~P.; de~Vries,~A.~H. {Comment on "On
  Using a Too Large Integration Time Step in Molecular Dynamics Simulations of
  Coarse-Grained Molecular Models"}. \emph{Phys. Chem. Chem. Phys.}
  \textbf{2010}, \emph{12}, 2254\relax
\mciteBstWouldAddEndPuncttrue
\mciteSetBstMidEndSepPunct{\mcitedefaultmidpunct}
{\mcitedefaultendpunct}{\mcitedefaultseppunct}\relax
\EndOfBibitem
\bibitem[van Gunsteren and Winger(2010)van Gunsteren, and Winger]{Marrink2010}
van Gunsteren,~W.~F.; Winger,~M. {Reply to the 'Comment on "On Using a Too
  Large Integration Time Step in Molecular Dynamics Simulations of
  Coarse-grained Molecular Models"'}. \emph{Phys. Chem. Chem. Phys.}
  \textbf{2010}, \emph{12}, 2254\relax
\mciteBstWouldAddEndPuncttrue
\mciteSetBstMidEndSepPunct{\mcitedefaultmidpunct}
{\mcitedefaultendpunct}{\mcitedefaultseppunct}\relax
\EndOfBibitem
\bibitem[Zwanzig(2001)]{Zwanzig2001}
Zwanzig,~R. \emph{{Nonequilibrium Statistical Mechanics}}; Oxford University
  Press: New York, 2001\relax
\mciteBstWouldAddEndPuncttrue
\mciteSetBstMidEndSepPunct{\mcitedefaultmidpunct}
{\mcitedefaultendpunct}{\mcitedefaultseppunct}\relax
\EndOfBibitem
\bibitem[Towns \latin{et~al.}(2014)Towns, Cockerill, Dahan, Foster, Gaither,
  Grimshaw, Hazlewood, Lathrop, Lifka, Peterson, Roskies, Scott, and
  Wilkins-Diehr]{xsede}
Towns,~J.; Cockerill,~T.; Dahan,~M.; Foster,~I.; Gaither,~K.; Grimshaw,~A.;
  Hazlewood,~V.; Lathrop,~S.; Lifka,~D.; Peterson,~G.~D.; Roskies,~R.;
  Scott,~J.~R.; Wilkins-Diehr,~N. {XSEDE: Accelerating Scientific Discovery}.
  \emph{Comput. Sci. Eng.} \textbf{2014}, \emph{16}, 62--74\relax
\mciteBstWouldAddEndPuncttrue
\mciteSetBstMidEndSepPunct{\mcitedefaultmidpunct}
{\mcitedefaultendpunct}{\mcitedefaultseppunct}\relax
\EndOfBibitem
\bibitem[Grime and Voth(2014)Grime, and Voth]{Grime2014}
Grime,~J. M.~A.; Voth,~G.~A. {Highly Scalable and Memory Efficient
  Ultra-Coarse-Grained Molecular Dynamics Simulations}. \emph{J. Chem. Theory
  Comput.} \textbf{2014}, \emph{10}, 423--431\relax
\mciteBstWouldAddEndPuncttrue
\mciteSetBstMidEndSepPunct{\mcitedefaultmidpunct}
{\mcitedefaultendpunct}{\mcitedefaultseppunct}\relax
\EndOfBibitem
\bibitem[Br{\'{a}}zdov{\'{a}} and Bowler(2008)Br{\'{a}}zdov{\'{a}}, and
  Bowler]{Brazdova2008}
Br{\'{a}}zdov{\'{a}},~V.; Bowler,~D.~R. {Automatic Data Distribution and Load
  Balancing with Space-filling Curves: Implementation in CONQUEST}. \emph{J.
  Phys. Condens. Matt.} \textbf{2008}, \emph{20}, 275223\relax
\mciteBstWouldAddEndPuncttrue
\mciteSetBstMidEndSepPunct{\mcitedefaultmidpunct}
{\mcitedefaultendpunct}{\mcitedefaultseppunct}\relax
\EndOfBibitem
\end{thebibliography}
\end{document}